\documentclass[%
 reprint,
superscriptaddress,
 amsmath,amssymb,
 aps,prl,
floatfix,
]{revtex4-1}

\usepackage{graphicx}
\usepackage[outdir=./]{epstopdf}
\usepackage{dcolumn}
\usepackage{bm}
\usepackage{mathtools}
\usepackage{color}
\usepackage{xspace}

\begin{document}

\preprint{APS/123-QED}
\title
{Impact of global structure on diffusive exploration of organelle networks}

\author{Aidan I. Brown}
\affiliation{Department of Physics, University of California, San Diego, San Diego, California 92093}

\author{Laura M. Westrate}
\affiliation{Department of Chemistry and Biochemistry, Calvin College, Grand Rapids, Michigan 49546}
\affiliation{Department of Molecular, Cellular and Developmental Biology, University of Colorado at Boulder, Boulder, Colorado 80309}

\author{Elena F. Koslover}
\email{ekoslover@ucsd.edu}
\affiliation{Department of Physics, University of California, San Diego, San Diego, California 92093}

\begin{abstract}	
We investigate diffusive search on planar networks, motivated by tubular organelle networks in cell biology that contain molecules searching for reaction partners and binding sites. Exact calculation of the diffusive mean first-passage time on a spatial network is used to characterize the typical search time as a function of network connectivity. We find that global structural properties --- the total edge length and number of loops ---  are sufficient to largely determine network exploration times for a variety of both synthetic planar networks and organelle morphologies extracted from living cells. For synthetic networks on a lattice, we predict the search time dependence on these global structural parameters by connecting with percolation theory, providing a bridge from irregular real-world networks to a simpler physical model. The dependence of search time on global network structural properties suggests that network architecture can be designed for efficient search without controlling the precise arrangement of connections. Specifically, increasing the number of loops substantially decreases search times, pointing to a potential physical mechanism for regulating reaction rates within organelle network structures.
\end{abstract}

\maketitle


Network models have been employed to describe and understand a wide variety of phenomena~\cite{boccaletti06}, ranging from transparently physical processes such as flow~\cite{durand07} and conductivity~\cite{cheianov07} to the more abstract examples of human physiology~\cite{bashan2012network,ivanov2016focus}, social interactions~\cite{wasserman94}, and mortality~\cite{farrell18}. Random walks on networks~\cite{masuda17} can model the dynamics of epidemic spreading~\cite{pu15}, animal foraging~\cite{perna14}, brain signaling~\cite{chavez10}, and electron transport~\cite{nelson99}. The rate at which such random walkers find target sites within the networks is known to depend on factors such as dimensionality~\cite{burioni05}, target connectivity~\cite{hwang12}, and number of shortest paths passing through the target~\cite{noh04}.

In comparison to generalized complex networks, spatial networks have physical constraints that limit connections to spatially proximal nodes~\cite{barthelemy11}. In addition, physical diffusion along network edges gives rise to broadly distributed non-exponential waiting times that depend on edge length~\cite{redner01}, in contrast to classic models of random walks on networks~\cite{noh04,barthelemy11,masuda17}. Thus, diffusive search on spatial networks is perhaps better described by physical variables such as fractal dimensionality~\cite{ben2000diffusion,benichou10} or tortuosity~\cite{shen2007critical}, rather than the number of nodes and edges often used to characterize general network structures.
 Random walks on spatial networks have similar dynamical properties to those in complex, porous, or crowded media~\cite{havlin87,condamin07,benichou14}. Although there has been progress in calculating how geometry affects diffusive search times on complex domains~\cite{condamin07,benichou10,benichou14}, and understanding the impact of distinct search strategies~\cite{chupeau15} there is little guidance on how to structure spatial networks to accelerate diffusive search.

Intracellular structures provide a key example where diffusive processes over complex geometries have an important role to play in cellular function. Reticulated organelles, such as the peripheral endoplasmic reticulum (ER)~\cite{westrate15,schwarz16} and mitochondria~\cite{collins02,rafelski12}, are composed of membranous tubules enclosing a single connected luminal volume, whose physical structure has recently been mapped in extensive detail~\cite{speckner18,harwig18}. These organelles constitute spatial networks that span throughout the cell interior, comprising hundreds of nodes and edges connected in a highly looped architecture~\cite{lin14,viana19}. Within these networks, proteins and other molecules diffuse to find reaction partners and binding targets. For instance, secretory proteins must encounter an exit site in order to leave the ER~\cite{hughes09} and DNA-binding proteins must find mitochondrial nucleoids to participate in DNA maintenance and replication~\cite{ruhanen10}.

The structure of these living networks is heavily regulated and likely functionally important~\cite{english13,viana19}. Mitochondrial network structure changes~\cite{rube04} during the cell cycle~\cite{margineantu02}, differentiation~\cite{shin16}, and disease~\cite{willems09}, suggesting mitochondrial morphology plays a role in physiological functions such as ATP production~\cite{chiaradonna06,ghosh18}. ER structure varies with cell specialization~\cite{schwarz16} and with mutations in morphogenic proteins associated with human pathologies~\cite{chen13,westrate15}.
Prior work analyzed the structure and morphogenesis of mitochondrial~\cite{viana19,zamponi18} and ER networks~\cite{lin14}, and mapped some basic parameters of molecular diffusion within these networks~\cite{dayel99,holcman18,viana19}. However the connection between network morphology and search efficiency has not been systematically addressed.

We investigate diffusive search on two network types. Firstly, we construct a variety of synthetic planar networks, with nodes on a lattice or homogeneously scattered and connected into a single component with varying arrangements of edges. Secondly, we use spatial networks extracted from imaging of yeast mitochondria and mammalian ER.
We find that typical search times on these biological structures are largely predicted by simple global structural parameters: the total edge length and loop number, which encompass network density and connectivity.

\section{Model}

To explore search efficiency, we analytically calculate the diffusive mean first-passage time (MFPT)~\cite{redner01} between an initial and a target node, given the connectivity and physical length of the network edges. Particle diffusion between nodes is governed by the propagator $G_{ij}(t)$, which gives the probability that a particle starting at node $i$ will be at node $j$ after time $t$, without passing through the target node. $\sum_j G_{ij}$ is the probability that the particle has never reached a target node. The MFPT to reach the target node $k$ is
\begin{equation}
	\label{eq:mfpt1}
\left\langle T_{ik} \right\rangle = \int_0^{\infty} t\left[-\frac{\partial}{\partial t}\sum_{j\neq k}G_{ij}(t)\right]dt = \sum_{j\neq k} \hat{G}_{ij}(s=0) \ .
\end{equation}
where  $\hat{G}_{ij}(s)$ is the Laplace-transform of $G_{ij}(t)$. Adapting recent work~\cite{koslover12}, the propagator is given by
\begin{equation}
\label{eq:similar}
\hat{G}_{ij}(s) = [(\boldsymbol{I}-\boldsymbol{\hat{P}})^{-1}]_{ij}\hat{Q}_j \ ,
\end{equation}
where $\boldsymbol{I}$ is the identity matrix, $\hat{P}_{nm}$ is the Laplace-transform of the flux of particles from node $n$ directly to a connected node $m$ without any intervening steps to other nodes, and $\hat{Q}_n$ is the Laplace-transformed probability that a particle starting at node $n$ has not arrived at another node. Paths that reach the target are assumed to leave the network entirely, so $\hat{P}_{nk}=0$. Eq.~\ref{eq:similar} generalizes earlier work~\cite{masuda17} to networks with distinct, non-exponential distributions for diffusion time along each edge. It differs from first-passage time calculations which assume all node-node transitions correspond to identical time steps~\cite{burioni05,maier17} or with infinitesimal time spent on edges~\cite{benichou14}, and from the numerical integration previously used to evaluate diffusion on systems of containers connected with tubes~\cite{lizana05}.

Inserting Eq.~\ref{eq:similar} into Eq.~\ref{eq:mfpt1} gives the MFPT between source node $i$ and target node $k$.
The elements 
$\hat{P}_{nm}(s=0)$ correspond to the probability that a particle starting at node $n$ will next step to node $m$, which depends only on the lengths $\ell_{nm}$ of the connecting edges:
\begin{equation}
\label{eq:pgen}
\hat{P}_{nm}(s=0) = \frac{\ell_{nm}^{-1}}{\sum_{w=1}^{\text{deg}(n)} \ell_{nw}^{-1}} \ ,
\end{equation}
where node $m$ and nodes $w$ are directly connected to node $n$.
Similarly, $\hat{Q}_{n}(s=0)$ gives the mean first passage time for a particle to arrive at any of the directly connected nodes from node $n$, with
\begin{equation}
\label{eq:qgen}
\hat{Q}_n(s=0) = \frac{1}{2D}\frac{\sum_{w=1}^{\text{deg}(n)}\ell_{nm}}{\sum_{w=1}^{\text{deg}(n)}\ell_{nm}^{-1}} \ ,
\end{equation}
where $D$ is the particle diffusivity and nodes $w$ are directly connected to node $n$ (derivations in Methods).

\section{Results}

\subsection{Visualizing mean first-passage times}

\begin{figure}[tbp] 
	\centering
	\hspace{-0.0in}
	\includegraphics[width=3.25in]{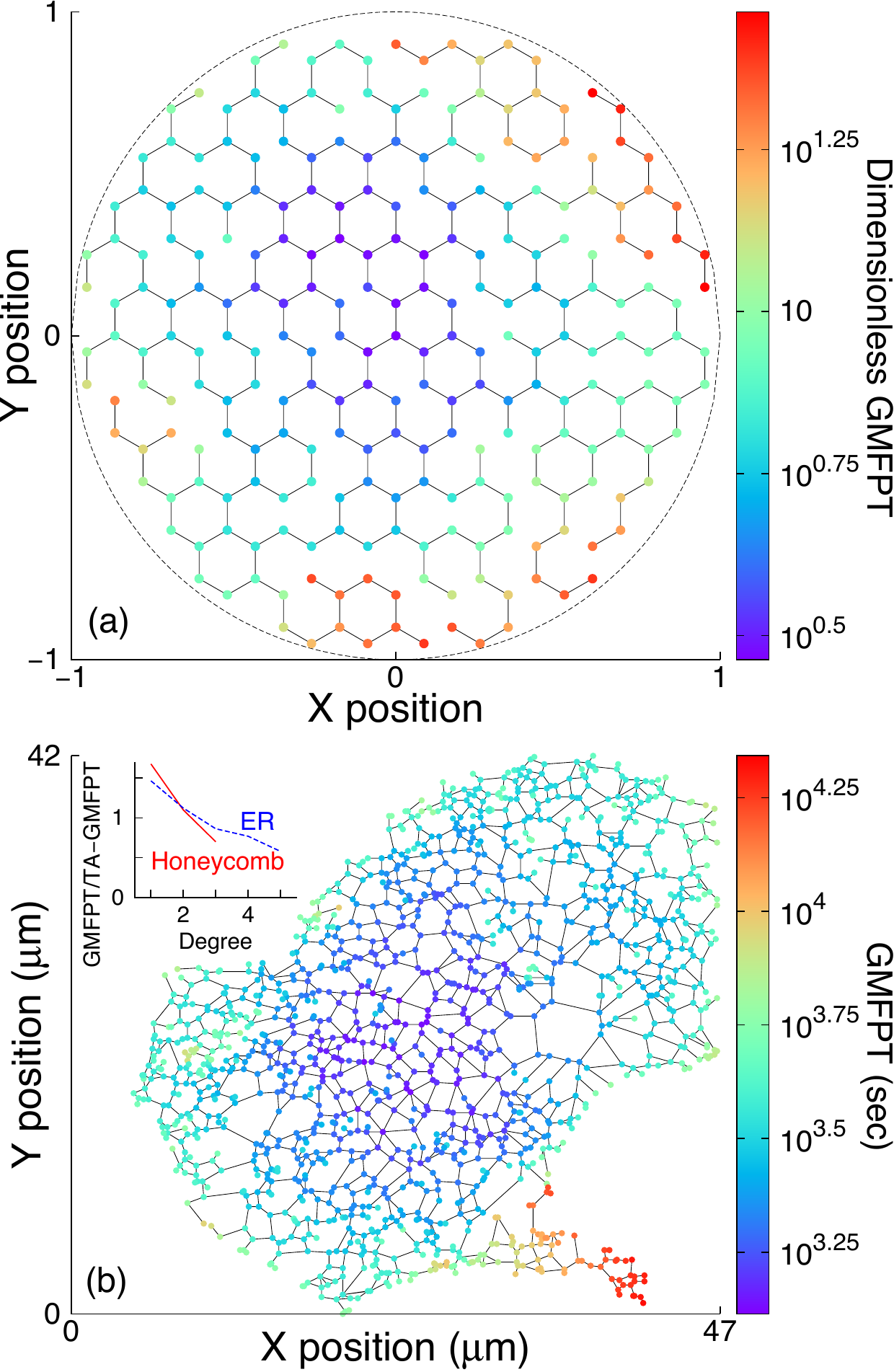}
	\caption{\label{fig:visual}
    {\bf Spatial variation of global mean first-passage time.} 
    Global mean first-passage time (GMFPT) represented by node color for an example (a) decimated honeycomb and (b) endoplasmic reticulum network. Times in (a) are nondimensionalized by $R^2/D$, where $R$ is the domain radius and $D$ is the particle diffusivity; times in (b) are given for a particle diffusivity of $D=1\mu\text{m}^2/\text{s}$.
    Inset of (b) shows GMFPT vs.\ node degree for both networks, normalized by the overall target-averaged GMFPT (TA-GMFPT). }
\end{figure}

Global mean first-passage time (GMFPT) is defined as the MFPT to a single target node averaged over all possible source nodes~\cite{tejedor09}. Figure~\ref{fig:visual}a shows GMFPTs for a `decimated' honeycomb network: a complete honeycomb network is constructed inside a circle of unit radius, and edges are removed while maintaining a single connected component. The radius of the circular domain $R$ sets the length-scale of the system, and the particle diffusivity $D$ sets the time-scale. Here, and in all subsequent results with synthetic networks, the first passage times are nondimensionalized by $R^2/D$.

Figure~\ref{fig:visual}b shows GMFPTs for a particle diffusing in an example ER network from a COS-7 cell (see Methods). Both networks in Fig.~\ref{fig:visual} have higher GMFPT for nodes nearer the network periphery, as compared to centrally located nodes. Better-connected (higher degree) nodes are found more quickly (Fig.~\ref{fig:visual}b inset).

\subsection{Loops and total edge length constrain TA-GMFPT}

Given the substantial GMFPT variation between target nodes on each network, we define a single metric characterizing the efficiency of target search processes on a particular network. Namely, the target-averaged GMFPT (TA-GMFPT) is defined as the GMFPT averaged over all possible target nodes in the network, and we use the TA-GMFPT as a typical `search time' hereafter.

\begin{figure*}[tbp] 
	\centering
	\hspace{-0.0in}
	\includegraphics[width=6.9in]{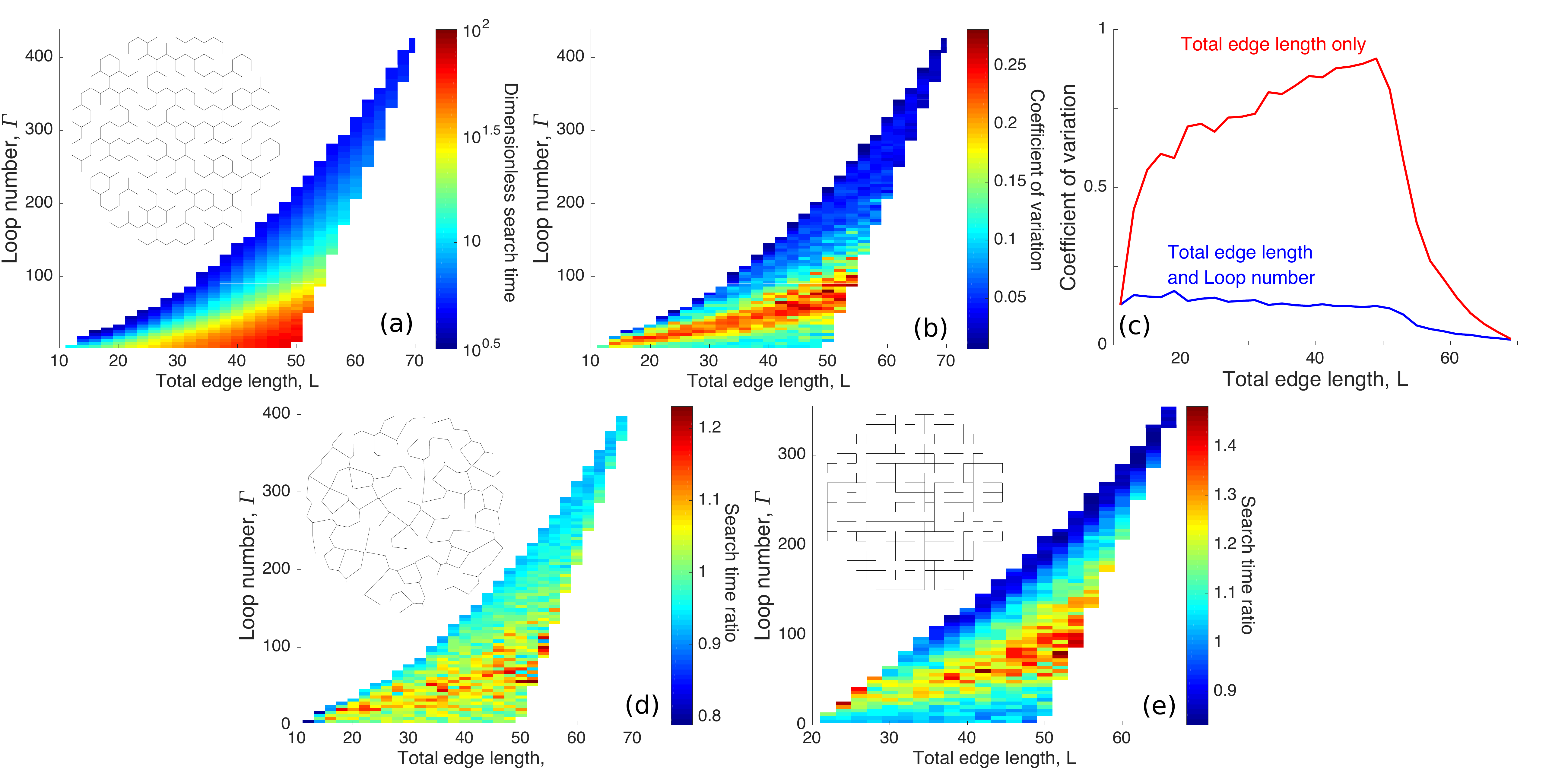}
	\caption{\label{fig:hexloopsedges}
		{\bf Geometric characteristics control typical search time.}
		(a) Target-averaged global mean first-passage time (TA-GMFPT) for honeycomb networks. Each network is sorted into a total edge length and loop number bin and mean TA-GMFPT is shown for each bin. 
		(b) TA-GMFPT coefficient of variation $c_{\text{v}}$ for each bin in (a).
		(c) The mean of the $c_{\text{v}}$ in each loop number bin in (b) at a given total edge length bin (Total edge length and Loop number, blue) and the $c_{\text{v}}$ across all loop numbers for a given total edge length bin (Total edge length only, red) --- details described in Methods.
		(d) Average TA-GMFPT ratio between Voronoi networks and the honeycomb networks in (a), and (e) between square networks and the honeycomb networks in (a).}
\end{figure*}

We investigate the impact of network structure on diffusive search over planar networks with nodes placed homogeneously throughout a circular domain. Decimated honeycomb networks are generated with different node densities and numbers of edges randomly removed, keeping only networks with all nodes connected. These networks all have the same spatial extent (set by domain radius $R=1$), but different connectivities and node densities.

The choice of decimated lattice planar network structures, with homogeneously distributed nodes, is motivated by suggestions that yeast mitochondrial networks are evenly spread along the cell surface~\cite{viana19} and ER networks in several adherent cell types span throughout the relatively flat periphery of the cell~\cite{lin14,friedman2011er}. We choose honeycomb networks because their three-way junction structure matches ER~\cite{shemesh14} and mitochondrial~\cite{viana19} networks, and the $120^{\circ}$ angles between edges at network junctions match the peak angle for ER junctions~\cite{lin14}. This construction enables the generation of a varied family of planar networks that connect well-distributed nodes while retaining some of the geometric and topological features of cellular network structures.

Each network is characterized by the sum of all edge lengths (`total edge length') $L$ and the cyclomatic number~\cite{barthelemy11}, which is the number of elementary cycles in the network, hereafter termed `loop number'. Loop number is given by $\Gamma = N_{\text{e}} - N_{\text{n}} + 1$, with $N_{\text{e}}$ the number of edges and $N_{\text{n}}$ the number of nodes. We use the loop number as a simple measure of redundant connectivity.

Figure~\ref{fig:hexloopsedges}a shows mean TA-GMFPT vs.\ total edge length $L$ and loop number $\Gamma$, averaged over many distinct decimated honeycomb networks. Larger $L$ increases search time, by increasing the one-dimensional volume of the search space. Higher $\Gamma$ substantially decreases search time --- for some $L$ values the mean search time varies by more than an order of magnitude over the explored range of $\Gamma$.

The TA-GMFPT coefficient of variation $c_{\text{v}}$ (ratio of standard deviation to mean) for a given total edge length $L$ and loop number $\Gamma$ does not exceed 0.3, with typical $c_{\text{v}}$ substantially lower (Fig.~\ref{fig:hexloopsedges}b). For most $L$ values, the $c_{\text{v}}$ given both $L$ and $\Gamma$ is significantly smaller than the $c_{\text{v}}$ given $L$ alone (Fig.~\ref{fig:hexloopsedges}c), demonstrating that both total edge length and loop number are necessary to accurately predict search times on a decimated lattice network. These parameters ($L$ and $\Gamma$) incorporate the number of nodes and edges on a network within a fixed spatial region, in a manner that highlights the network density and connectivity, respectively.

To establish the utility of total edge length $L$ and loop number $\Gamma$ in predicting network search times, we consider several alternate network architectures, each averaged over many distinct individual networks. Decimated Voronoi networks are generated by randomly placing points within a circle with an exclusion radius around each preceding point, constructing a Voronoi tessellation using these points, and removing edges while maintaining all nodes in the single connected component. These decimated Voronoi networks maintain the three-way junction geometry of honeycomb networks, and their mean search times are very similar to decimated honeycomb networks with the same $L$ and $\Gamma$ (Fig.~\ref{fig:hexloopsedges}d). We also generate decimated square networks, which have node degrees up to 4. The ratio of search times between these square networks and decimated honeycomb networks, matched by $L$ and $\Gamma$, shows greater variation (Fig.~\ref{fig:hexloopsedges}e). Nonetheless, search times on these square networks are generally within $40\%$ of comparable honeycomb networks -- this deviation is small in comparison to the orders of magnitude variation in the TA-GMFPT over the network structures in Fig.~\ref{fig:hexloopsedges}a. Figure~\ref{fig:hexloopsedges} thus highlights the importance of total edge length and loop number in determining diffusive search times over a broad variety of planar network structures with well-distributed nodes.

\subsection{Search on cell biology networks}

\begin{figure*}[tbp] 
	\centering
	\hspace{-0.0in}
	\includegraphics[width=6.9in]{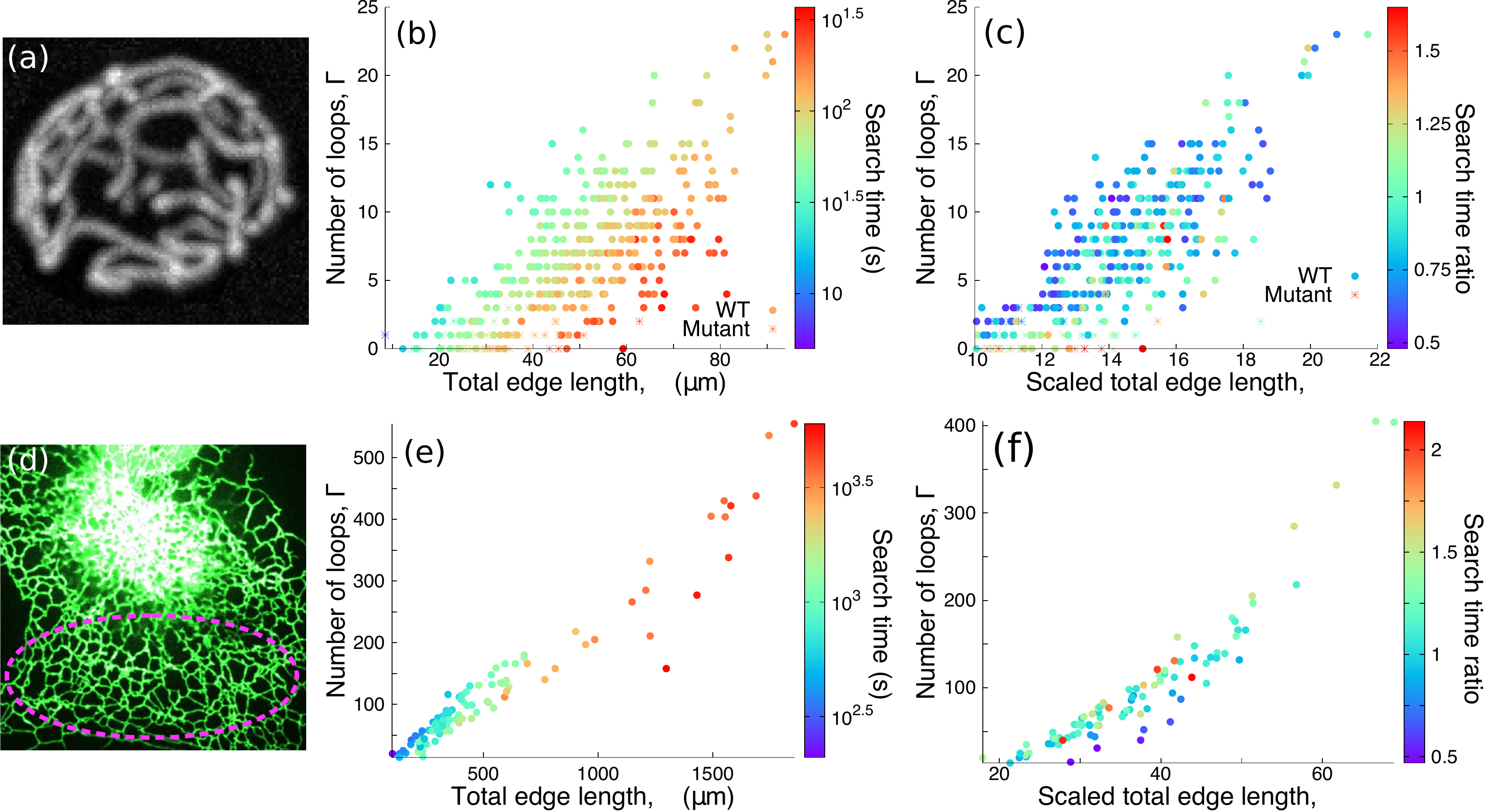}
	\caption{\label{fig:compareloopsedges}
    {\bf Search on cell biology networks is described by geometric characteristics.}
    (a) Fluorescence image of yeast mitochondrial network~\cite{viana19}.
    (b) Target-averaged global mean first-passage time (TA-GMFPT) on mitochondrial networks~\cite{viana19}, for particles with diffusivity $D=1\mu\text{m}^2/\text{s}$.
    (c) Ratio between TA-GMFPTs on mitochondrial networks, scaled to unit spatial extent, and the expected values at equivalent loop number and  total edge length on honeycomb networks in Fig.~\ref{fig:hexloopsedges}a. Only mitochondrial networks with scaled total edge length and loop number that overlap honeycomb networks are shown.
    (d) Fluorescence image of endoplasmic reticulum (ER) in COS-7 cell, with a region of peripheral ER network indicated.
    (e) TA-GMFPT on ER networks, for particles with diffusivity $D=1\mu\text{m}^2/\text{s}$.
    (f) Similar ratio to (c) between TA-GMFPTs on scaled ER and honeycomb networks.}
\end{figure*}

We also analyze search times on intracellular reticulated organelle network structures. Figure~\ref{fig:compareloopsedges}a shows an example fluorescent image of a yeast mitochondrial network~\cite{viana19}. Figure~\ref{fig:compareloopsedges}b shows TA-GMFPT from 350 mitochondrial networks~\cite{viana19}, 
which exhibit features similar to the honeycomb networks in Fig.~\ref{fig:hexloopsedges}a: approximate prediction of TA-GMFPT by total edge length $L$ and loop number $\Gamma$, and a substantial decrease in search time as $\Gamma$ increases and $L$ decreases. Mitochondrial networks from wild-type cells and mutant cells with mitochondrial fission and fusion proteins knocked out occupy distinct regions of the $\Gamma$ vs.\ $L$ plane. However, for given values of these two structural parameters, the two network types (wild-type and mutant) exhibit similar search times. The fluorescent image in Figure~\ref{fig:compareloopsedges}d shows an example ER network. We calculated the TA-GMFPT for regions of 103 such peripheral ER networks (Figure~\ref{fig:compareloopsedges}e). Although ER network structures are restricted to relatively high looping number for each total edge length, the search times vary similarly to honeycomb networks (Fig.~\ref{fig:hexloopsedges}a) and mitochondrial networks (Fig.~\ref{fig:compareloopsedges}a).

While the synthetic honeycomb, Voronoi, and square networks in Fig.~\ref{fig:hexloopsedges} are planar, mitochondrial and ER networks exist in three-dimensional intracellular volumes. However, imaging of peripheral ER networks in COS7 cells indicates that these structures are relatively flat, with rarely observed crossing of tubules outside the typical 3-way junction nodes~\cite{friedman2011er,chen13,holcman18}.
Three dimensional deformation in the paths of individual edges would reduce the overall effective diffusion coefficient in a planar projection~\cite{adler2019conventional}, but would not substantially alter the global search trends described here. Mitochondrial networks in budding yeast cells tend to remain at the cell surface, with little incursion into the three-dimensional bulk of the cell~\cite{viana19}. These networks are thus essentially confined to a two-dimensional manifold in the shape of a spherical shell. For simplicity we also approximate them as planar, neglecting the large-scale curvature of the spherical surface. 


Investigation of how network structural characteristics facilitate searches must account for the expected search time increase as domain size increases. To compare the mitochondrial and ER networks to idealized lattice-like structures, the cellular networks are scaled to the same physical area as the unit circle containing honeycomb networks. Three-dimensional mitochondrial network coordinates extracted from imaging are projected onto a spherical surface. The network area is estimated using a convex hull of both ER nodes on the plane and projected mitochondrial nodes on the sphere  (details in Methods).
 This allows comparison of search times between networks with the same spatial extent but different node density and connectivity.
 
  Figures~\ref{fig:compareloopsedges}c and \ref{fig:compareloopsedges}f plot the ratio of each mitochondrial and ER network search time, respectively, to the honeycomb network search time at the corresponding total edge length and loop number. The ratios for both mitochondrial and ER networks are near unity, suggesting these organelle network structures have similar search characteristics and dependence on total edge length and loop number as synthetic lattice-like networks.

\subsection{Dependence of Search Times on Network Morphology}

\begin{figure*}[tbp] 
	\centering
	\hspace{-0.0in}
	\includegraphics[width=6.9in]{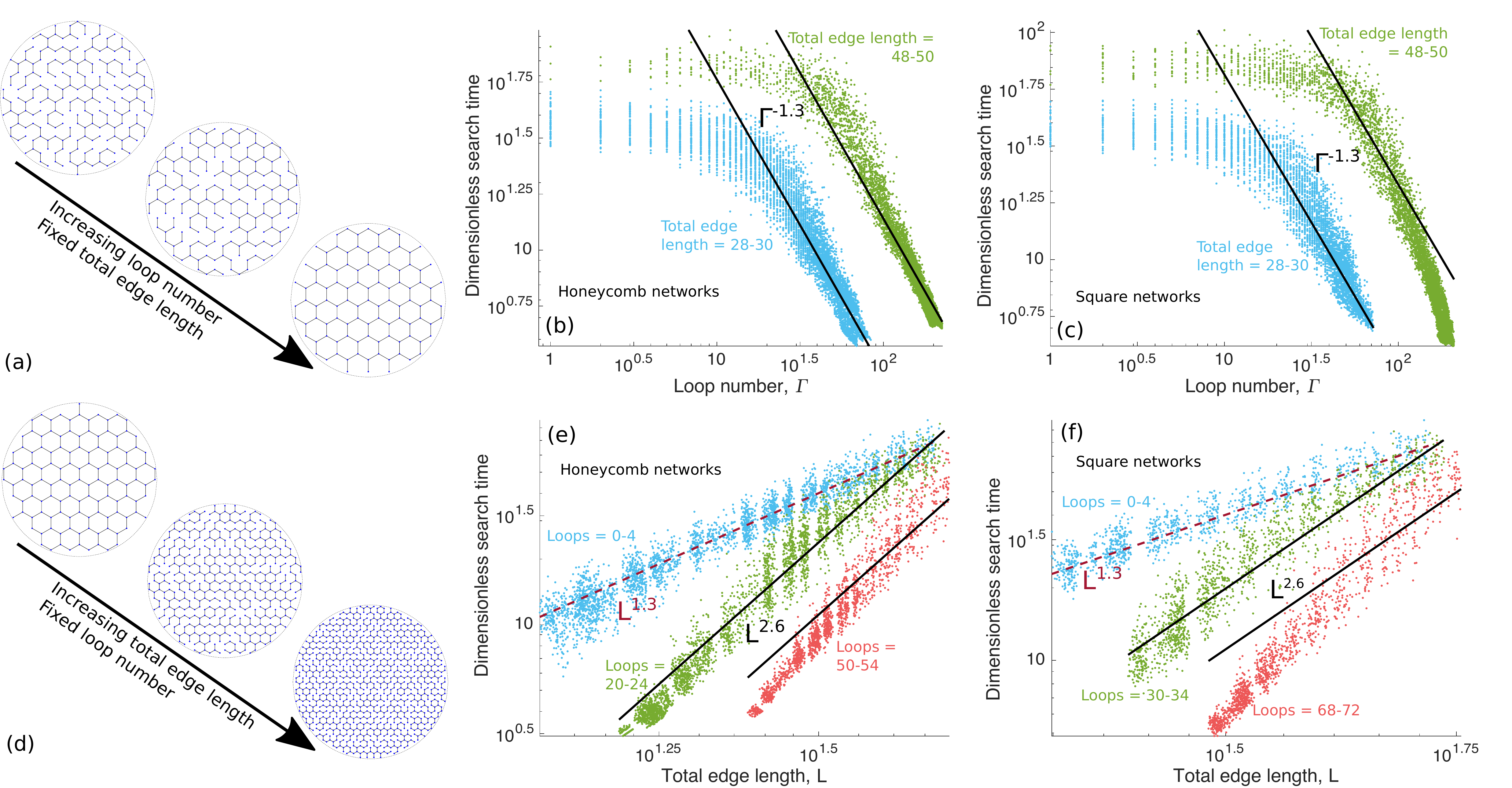}
	\caption{\label{fig:loopsedgesonedim}
    {\bf Variation of search time with geometric quantities.}
    (a) Example honeycomb networks showing increasing loop number while total edge length is held fixed.
    Using target-averaged global mean first-passage time (TA-GMFPT) from honeycomb networks in Fig.~\ref{fig:hexloopsedges}a, TA-GMFPT vs.\ loop number $\Gamma$ for fixed total edge length range for (b) honeycomb networks from Fig.~\ref{fig:hexloopsedges}a and (c) square networks from Fig.~\ref{fig:hexloopsedges}d. Each colored point indicates the search time and loop number for an individual network, with different colored points indicating a different total edge length range. Solid black lines show $\Gamma^{-1.3}$ power laws predicted from percolation theory.
    (d) Example honeycomb networks showing increasing total edge length while loop number is held fixed.
    Similar to (b) and (c), TA-GMFPT vs.\ total edge length $L$ for fixed loop number range for (e) honeycomb and (f) square networks. Each colored point indicates the search time and total edge length for an individual network, with different color points indicating a different loop number range. Dashed red lines show $L^{1.3}$ power laws, and solid black lines show $L^{2.6}$ power laws predicted from percolation theory. Prefactors for all power-laws were selected arbitrarily to serve as a guide to the eye.}
\end{figure*}

Using decimated honeycomb and square lattice network structures (Fig.~\ref{fig:hexloopsedges}), we explore the dependence of search time on total edge length $L$ and loop number $\Gamma$ (Fig.~\ref{fig:loopsedgesonedim}). Increasing loop number at a constant total edge length corresponds to networks with less dense nodes that are more completely connected (Fig.~\ref{fig:loopsedgesonedim}a). 
The search time varies distinctly for low vs.\ high loop numbers $\Gamma$. Search time weakly depends on $\Gamma$ for low $\Gamma$ (Fig.~\ref{fig:loopsedgesonedim}b,c), indicating that adding a few loops, within a primarily tree-like structure, will not substantially affect the search process. In contrast, search time steeply decreases with rising $\Gamma$ for higher $\Gamma$ values (Fig.~\ref{fig:loopsedgesonedim}b,c). This suggests that once a threshold number of loops is reached, further added loops can significantly decrease search time. High search time variability in Fig.~\ref{fig:hexloopsedges}b,d,e aligns with the neighborhood of these thresholds in Fig.~\ref{fig:loopsedgesonedim}b,c, suggesting that at the threshold where loop number begins to perturb global transport, the precise arrangement of the loops can have a substantial impact on search time. 

By contrast, increasing total edge length $L$ at a constant loop number corresponds to denser, less well connected networks (Fig.~\ref{fig:loopsedgesonedim}d). The dependence of search time on total edge length $L$ becomes more steep as loop number increases (Fig.~\ref{fig:loopsedgesonedim}e,f).

The decimated honeycomb and square lattice networks resemble percolation systems, where the fraction of bonds retained is above the critical percolation value $p_\text{c}$. 
Random walks in such systems are effectively diffusive above a certain correlation length, and for planar networks should have the same scaling properties as two-dimensional diffusion~\cite{gefen83,stauffer94}. In particular, we expect the search time to be largely independent of node density and to scale as  $T \sim D^{-1}$, where $D$ is the effective diffusivity (see Methods for details). Near the percolation threshold, $D\sim(p-p_{\text{c}})^{\mu}$, where $p$ is the fraction of lattice bonds remaining and $\mu\simeq1.30$ for two-dimensional lattices~\cite{stauffer94}.


By treating our synthetic networks as large clusters in a two-dimensional system approaching percolation, we derive the expected dependence of search times on total edge length and loop number (see Methods section). Namely, when loop number is very low ($\Gamma \ll L$), then search times are expected to be independent of $\Gamma$ but to scale with total edge length as $T \sim L^\mu$. Both of these expected relationships are consistent with the observed dependence of search times on $L$ and $\Gamma$ for synthetic networks with low loop numbers (Fig.~\ref{fig:loopsedgesonedim}). 
%

We note that the $L^{1.30}$ dependence is intermediate between two extreme cases of loopless networks within a fixed-area domain. One extreme includes linear structures that snake through the domain without branching, or comb-like networks with a single backbone connecting many individual branches, which both exhibit MFPTs scaling as $\sim L^2$~\cite{redner01}. The other extreme is self-similar tree-like networks~\cite{redner01,lin2010determining}, which have MFPTs that vary as $\sim L$ when scaled down to unit physical extent (see Methods).

For high loop numbers ($\Gamma \gg L$), search times in a cluster close to the percolation transition are expected to depend on both total edge length and loop number as $T\sim L^{2\mu}\Gamma^{-\mu}$ (see Methods). In Fig.~\ref{fig:loopsedgesonedim}b,c, this $T\sim\Gamma^{-1.30}$ dependence is consistent with search times for low total edge lengths, but does not entirely explain search time behavior at the highest total edge lengths and high loop numbers. Similarly, in Fig.~\ref{fig:loopsedgesonedim}e,f, the predicted $T\sim L^{2.60}$ dependence is consistent with search times for intermediate loop numbers, but does not entirely explain search time behavior for the highest loop numbers. The transition from intermediate connectivity to poor connectivity ($\Gamma \lesssim L$) is evident in the top right corner of Fig.~\ref{fig:loopsedgesonedim}e,f, where the search time dependence on total edge length begins to shift from $T\sim L^{2\mu}$ to $T\sim L^\mu$. We note that networks with the lowest total edge length for a given loop number (or highest loop number for a given total edge length) correspond to the most fully connected lattices, which are far from the percolation transition. We would thus expect the aforementioned scaling relationships to break down in this regime, as is seen for the data points with highest $L$ and $\Gamma$ in Fig.~\ref{fig:loopsedgesonedim}c, and for the lowest $L$, highest $\Gamma$ in Fig. ~\ref{fig:loopsedgesonedim}e,f.

\section{Discussion}

We have investigated the characteristics that control diffusive search time on planar networks connecting homogeneously distributed nodes over a compact domain. To this end, we employ an exact calculation of mean first-passage time on a spatial network (Eqs.~\ref{eq:mfpt1}-\ref{eq:qgen}) based on network connectivity and edge lengths.

Search times in a complex medium are known to depend not only on the spatial structure of the domain but also on the dynamic nature of the search process, with the dimensionality of the walk (defined by $\left<x^2\right> \sim  t^{2/d_{\text{w}}}$) determining a phase transition between compact and non-compact search processes~\cite{carretero2012phase,condamin07,benichou10,benichou14}. 
In the analysis presented here, changing the dimensionality for particle dynamics would alter how splitting probabilities and waiting times for node-to-node transitions depend on edge length (Eqs.~\ref{eq:pgen}, \ref{eq:qgen}), as well as likely modifying the scaling behavior of search times with network structure near the percolation transition (as discussed in Methods). Such extension to different dynamical processes is outside the scope of this work and is left as a fruitful area for further study.

Throughout our calculations, particle motion along network edges is assumed to be purely diffusive ($d_{\text{w}}=2$). This assumption of diffusive transport is consistent with past analyses of large-scale particle spreading on both ER and mitochondrial structures, as measured by fluorescence recovery after photobleaching~\cite{partikian1998rapid,dayel99,sbalzarini2005effects,dieteren2011solute}. More recent single-particle tracking studies in the ER indicate that membrane proteins move diffusively, while luminal proteins may in some cases be driven by random processive flows along the network edges~\cite{holcman18}. Subdiffusive behavior ($d_{\text{w}}>2$), commonly attributed to fractional Brownian motion in a viscoelastic medium, has been observed for a variety of intracellular particles of size comparable to organelles or RNA-protein complexes ($r\gtrsim 50$nm)~\cite{yamada2000mechanics,tolic2004anomalous,lampo2017cytoplasmic}. However, smaller particles such as individual proteins ($\sim 5$nm) often exhibit diffusion-like motion~\cite{etoc2018non}. These observations motivate our choice to focus on diffusive exploration for proteins in the ER and mitochondrial networks.


We assess typical search time on each network by averaging all combinations of source and target nodes. Diffusive search time on networks with homogeneously distributed nodes, including both synthetic networks and those from intracellular structures, is found to be largely predicted by simple geometric characteristics: total edge length and loop number (Figs.~\ref{fig:hexloopsedges}, \ref{fig:compareloopsedges}), which characterize network density and connectivity.
 Increasing loop number substantially decreases diffusive search time, while increasing total edge length can steeply increase the search time (Figs.~\ref{fig:hexloopsedges}-\ref{fig:loopsedgesonedim}). Search times on ER and mitochondrial networks are comparable to those computed for idealized planar lattice structures with equivalent loop number and edge length (Fig.~\ref{fig:compareloopsedges}c,f), emphasizing the sufficiency of these two global structural parameters for determining diffusive search efficiency on real-world networks.

Using percolation theory to predict diffusivity, which is inversely related to the search time, largely describes the dependence of search time on total edge length and loop number for networks constructed from decimated planar lattices. This link to percolation theory highlights the importance of network connectivity, in the form of the bond fraction $p$, for determining search times.  
Although the bond fraction is well-defined for idealized regular planar lattices, it is not a meaningful parameter for realistic cellular networks, such as ER and mitochondrial structures.
For these and other off-lattice networks, we show that the simply measurable parameters of total edge length and loop number can be used to predict diffusive search behavior. We have outlined how these network parameters can be connected to an effective bond fraction and thus to the wide range of results available for percolation systems.

Typical mitochondrial and ER networks have many loops, which accelerate search (Figs.~\ref{fig:compareloopsedges}b,e). These search-accelerating loops align with Murray's law for vasculature radius~\cite{sherman81,mcculloh03} or the balance of competing constraints in fungi~\cite{tero10,heaton12}, suggesting biology is capable of optimizing transport networks and that networks with many loops may have been partly selected for efficient diffusive transport.

Mitochondrial and endoplasmic reticulum networks extend through much of the cellular volume to interact and form direct contacts with other organelles~\cite{klecker2014making,phillips2016structure,valm2017applying}, providing connections between various subcellular systems. 
A well-connected reticulated network structure enables proteins within these organelles to explore many contact sites without requiring export or relying on the slower dynamics of the organelles themselves. 
 Mitochondrial and endoplasmic reticulum network morphology can vary based on cell specialization and the functional state of the organelle~\cite{benard2008ultrastructure,schuck2009membrane}, possibly modulating signals carried by diffusion through mitochondrial and endoplasmic reticulum networks. Our results indicate that higher loop numbers in the network would decrease diffusive search times and thereby may speed transmission of diffusive signals to organelle contact sites. This potential for organelle functional and signaling modulation through network structural properties is similar to the influence of topology and significance of time delay for organismal states described by network physiology~\cite{bashan2012network,bartsch2015network,ivanov2016focus}.

Our finding that loops speed diffusive search points towards key structural criteria for spatial networks whose function relies on efficient diffusive transport, including the intracellular networks studied here.
Earlier work on random walks in complex networks showed that tree networks maximize the TA-GMFPT (i.e., lead to the slowest search)~\cite{benichou14}, which suggested that some loops may lead to more efficient diffusive search.

By contrast to networks designed for diffusive transport, the optimal spatial network structure for potential-driven flow in a variety of scenarios is a loopless tree~\cite{banavar00,durand07,bohn07,bernot09,hu13}. However, loops can assist network flow-based transport outside of steady state, providing resiliency to damage and fluctuations~\cite{katifori10,corson10,hu13}. Mitochondrial~\cite{zamponi18} and ER networks~\cite{speckner18} are very dynamic, and resiliency to edge removal may be another benefit of the many loops in these cell biology networks. Although we do not include edge or loop production cost~\cite{katifori10,corson10}, our analysis can establish the utility of these network components for improving transport.

The connection between network structure and diffusive distribution efficiency indicates a potential link between architecture and functionality for cellular organelles such as the ER and mitochondria. The dependence of search efficiency on global structural properties of the network suggests that cells may be able to regulate biochemical kinetics without precise local arrangement of network connections.

\section{Methods}

\subsection{Deriving P and Q}
Equations~\ref{eq:mfpt1} and \ref{eq:similar} give the MFPT between source node $i$ and target node $k$,
\begin{equation}
\label{eq:mfpt_final2}
\left\langle T_{ik}\right\rangle = \left[\left(\mathbf{I - \mathbf{\hat{P}}}\right)^{-1} \hat{\vec{Q}}\right]_{i,s=0} \ .
\end{equation}

To find $\hat{P}_{ij}$ and $\hat{Q}_j$, we consider, as an example, a particle at a degree-three node with edges of length $\ell_1 \leq \ell_2 \leq \ell_3$ connecting to other nodes. Trajectories that reach node 1 before nodes 2 or 3 can be constructed from excursions a distance $\ell_1$ from the initial node, with first-passage time distribution
\begin{align}
	\label{eq:excursions}
	&P_1(t) = \frac{1}{3}f_{2\ell_1}(t) + \int_0^t dt_2 \int_0^{t_2} dt_1 \nonumber \\
	& \frac{1}{3}f_{2\ell_1}(t-t_2)F_{\ell_2}(t_2-t_1)\frac{1}{3}f_{2\ell_1}(t_1) + \ldots \ .
\end{align}
For a diffusing particle that starts at position $\ell_1$ on an interval with absorbing boundaries at $x=0$ and $x=d$, the function $f_d(t)$ gives the total flux out of the interval at time $t$ and the function $F_d(t)$ gives the flux at the $x=0$ boundary.   
The first term of Eq.~\ref{eq:excursions} represents a trajectory that first reaches a distance $\ell_1$ from the initial node when it arrives at node 1. The second term of Eq.~\ref{eq:excursions} is for a particle that reaches $\ell_1$ from the initial node along the edge to node 2, returns to the initial node without first reaching node 2, and then diffuses to node 1. A Laplace transform $t\rightarrow s$ converts the convolutions over sequential steps into products, giving 
\begin{subequations}
	\begin{align}
		\hat{P}_1 &= \frac{1}{3}\hat{f}_{2\ell_1} + \frac{1}{3}\hat{f}_{2\ell_1}(\hat{F}_{\ell_2}+\hat{F}_{\ell_3})\bigg[\frac{1}{3}\hat{f}_{2\ell_1}  \nonumber \\
		&+ \frac{1}{3}\hat{f}_{2\ell_1}(\hat{F}_{\ell_2}+\hat{F}_{\ell_3})\left[\frac{1}{3}\hat{f}_{2\ell_1} + \ldots \right]\bigg] \\
		& = \frac{1}{3}\hat{f}_{2\ell_1}\sum_{n=0}^{\infty}\left[\frac{1}{3}\hat{f}_{2\ell_1}(\hat{F}_{\ell_2} + \hat{F}_{\ell_3})\right]^n \\
		\label{eq:geometric}& = \frac{1}{3}\hat{f}_{2\ell_1}\bigg/\left[1 - \frac{1}{3}\hat{f}_{2\ell_1}(\hat{F}_{\ell_2} + \hat{F}_{\ell_3})\right] \ ,
	\end{align}
\end{subequations}
with the last line from the infinite summation of a geometric series.

The Laplace-transforms of $f_{d}(t)$ and $F_{d}(t)$ are~\cite{redner01}
\begin{subequations}
	\label{eq:onedfpts}
	\begin{equation}
	\hat{f}_{d}(s) = 2\sinh\left[\sqrt{s/D}(d-\ell_1)\right]\Big/\left[\sinh(\sqrt{s/D}d)\right] \ ,
	\end{equation}
	\begin{equation}
	\hat{F}_{d}(s) = \sinh\left[\sqrt{s/D}(d-\ell_1)\right]\Big/\left[\sinh(\sqrt{s/D}d)\right] \ ,
	\end{equation}
\end{subequations}
where $D$ is the particle diffusivity. Inserting Eq.~\ref{eq:onedfpts} into Eq.~\ref{eq:geometric} and taking $s\to0$,
\begin{equation}
\hat{P}_1(s=0) = \ell_2\ell_3/\left(\ell_1\ell_2 + \ell_2\ell_3 + \ell_1\ell_3\right) \ .
\end{equation}
\begin{equation}
\label{eq:qfirst}
Q_j(t) = 1 - \sum_{w=1}^{3} P_{jw}(t) \ ,
\end{equation}
where the sum is over the nodes directly connected to node $j$.
\begin{equation}
\label{eq:qsecond}
\hat{Q}_j(s) = \frac{1}{s}\left[1 - \sum_1^{3} \hat{P}_{jw}(s)\right] \ .
\end{equation}
Expanding for small $s$ gives,
\begin{equation}
\begin{split}
\hat{Q}_j(s=0) & = -\frac{\partial}{\partial s}\sum_1^{3}\hat{P}_{jw}(s=0)  \\
& = \frac{1}{2D}\frac{\ell_1 + \ell_2 + \ell_3}{\ell_1^{-1} + \ell_{2}^{-1} + \ell_3^{-1}} \ .
\end{split}
\end{equation}
More generally, $\hat{P}_{ij}$ and $\hat{Q}_j$ are
\begin{equation}
\label{eq:pgen2}
\hat{P}_{nm}(s=0) = \frac{\ell_{nm}^{-1}}{\sum_{w=1}^{\text{deg}(n)} \ell_{nw}^{-1}} \ ,
\end{equation}
\begin{equation}
\label{eq:qgen2}
\hat{Q}_n(s=0) = \frac{1}{2D}\frac{\sum_{w=1}^{\text{deg}(n)}\ell_{nm}}{\sum_{w=1}^{\text{deg}(n)}\ell_{nm}^{-1}} \ ,
\end{equation}
where node $m$ and nodes $w$ are directly connected to node $n$.

\subsection{Generating networks}
\subsubsection{Synthetic networks}
We generate `decimated' networks by constructing a complete network and removing a set of edges, subject to the condition that all initial nodes remain attached to all other nodes in a single connected component when edges are removed. Many network variations can be constructed from one complete network by varying the number and identity of removed edges.

For honeycomb and square networks, initial complete networks are constructed as a lattice within a circle of radius one, with nearest neighbors connected by an edge. The lattice size is varied to obtain complete networks with different node densities. $6.2\times10^4$ honeycomb networks are generated for data in Fig.~\ref{fig:hexloopsedges}a, $3.6\times10^4$ Voronoi networks for Fig.~\ref{fig:hexloopsedges}d, and $6.5\times10^4$ square networks for Fig.~\ref{fig:hexloopsedges}e.

To construct a Voronoi network, we first randomly place points within a circle of radius one, subject to the condition that each subsequent point cannot be within an exclusion radius of all preceding points. When nodes can no longer be placed (as the entire circle is blocked with the exclusion radius of at least one point), a Voronoi tesselation is constructed around these points. The boundaries of the Voronoi tesselation cells form the network. The exclusion radius around the initial points is varied to obtain networks with different node densities.

\subsubsection{Mitochondrial networks}
Spatial coordinates and network connections for 350 mitochondrial networks from \emph{Saccharomyces cerevisiae} budding yeast cells, obtained using Mitograph software, were generously provided by Matheus Viana and Susanne Rafelski~\cite{viana19}. The networks we analyze include wild-type cells 
and $\Delta dnm1\Delta fzo1$ mutant cells lacking proteins for mitochondrial fission and fusion. For each cell, we used the largest connected component.

Mitochondrial networks have relatively few nodes and edges in comparison to the synthetic networks. Nodes with degree two were added along edges to ensure individual edge lengths were approximately homogeneous, facilitating comparison with decimated lattice networks. Specifically, sufficient nodes were added to make all individual node-to-node edges shorter than the shortest full edge in the original network, and shorter than a $1\mu$m length ceiling. This procedure does not change the geometry or topology of the original network, but does redefine the set of target nodes used for the calculation of the TA-GMFPT. 

\subsubsection{Endoplasmic reticulum networks}
COS-7 cells were purchased from ATCC (Catalog $\#$ ATCC-CRL1651) and were grown in Dulbecco’s modified Eagle medium (DMEM) supplemented with 10$\%$ fetal bovine serum (FBS) and 1$\%$ penicillin/streptomycin (P/S). Prior to imaging experiments, COS-7 cells were seeded in a 6-well, plastic bottom dishes at $1\times10^5$ cells/mL about 18 hours prior to transfection. Plasmid transfections were performed as described previously~\cite{hoyer18}. For all imaging experiments, the ER was fluorescently labeled with 0.2 $\mu\text{g}$ KDEL venus transfected into each well of a 6-well dish~\cite{english13b}. Live cells were imaged at 37$^\circ$C in Fluorobrite imaging media (Invitrogen) supplemented with 10$\%$ FBS. Confocal Z-stack images of the peripheral ER were collected using Micromanager Imaging Software with a step size of 0.2 $\mu\text{m}$. All images were acquired on an inverted fluorescent microscope (TE-2000-U; Nikon) equipped with a Yokogawa spinning-disk confocal system (CSU-Xm2; Yokogawa CSU X1)~\cite{hoyer18}. Images were taken with a 100$\times$ NA 1.4 oil objective on an electron-multiplying charge-coupled device (CCD) camera 50$\times$50 (Andor). Images were acquired with Micromanager Imaging Software and then analyzed, merged and contrasted using Fiji (ImageJ)~\cite{schindelin12}.

A large continuous region of the peripheral endoplasmic reticulum network was selected from each image. The endoplasmic reticulum from this region of each image was skeletonized,
and node and edge data from the skeleton extracted, using Fiji (ImageJ). Node and edge data was analyzed to extract a network structure, assuming nodes within 0.001$\mu$m of one another are the same node. For each cell, the largest connected component was used. We obtained 103 ER networks.

\subsection{Coefficient of variation}

This section describes how Fig.~\ref{fig:hexloopsedges}c was obtained.

In Fig.~\ref{fig:hexloopsedges}b, network structures are sorted into bins according to their total edge length (bin size of 2) and loop number (bin size of 4). For each bin
there is a mean search time $m_{ij}$ (mean of the TA-GMFPTs for all networks falling into the bin) and a variance of the search time $\sigma_{ij}^2$, where  $i, j$ indicate the bin indices for total edge length and loop number, respectively.

In the Fig.~\ref{fig:hexloopsedges}c the red curve labeled `Total edge length only' is the coefficient of variation over all loop numbers given a total edge length bin $i$. This coefficient depends both on the variance within individual bins and the overall variability from bin to bin. It is given by
\begin{equation}
\begin{split}
c_{\text{v},i} & = \frac{1}{\left<m\right>_i}\sqrt{\left<\sigma^2\right>_i + \text{var}_i(m)}\ , \\
\end{split}
\end{equation}
where $\left<m\right>_i = \frac{1}{n_i} \sum_j m_{ij}$ is the search time averaged over all bins with a given edge length, $\left<\sigma^2\right>_i = \frac{1}{n_i} \sum_j \sigma_{ij}^2$ is the average of variance within each bin, and $\text{var}_i(m)  = \frac{1}{n_i} \sum_j m_{ij}^2 - \left<m\right>_i^2$ is the variance of mean search times across all bins for the given edge length. For each edge length, the averages are done over $n_i$ bins containing at least 10 networks.

In the Fig.~\ref{fig:hexloopsedges}c the blue curve labeled `Total edge length and Loop number' gives the average of the coefficients of variance for each individual bin fixing both edge length and loop number. The average is carried out over all bins corresponding to a particular total edge length:
\begin{equation}
c_{\text{v},i} = \frac{1}{n_i}\sum_j\frac{\sigma_{ij}}{m_{ij}} \ .
\end{equation}

\subsection{Network size scaling}
To make a direct comparison between search times for synthetic networks constrained to a circle of radius one, and search times for networks from cell biology, we scale lengths in the cellular networks such that the effective  area spanned by the network matches the synthetic network area of $\pi$ (circle of radius one). Search times are scaled by the length scaling factor squared, as diffusive processes in one dimension occur in a time proportional to length squared.

Three-dimensional points along the largest connected component of each mitochondrial network skeleton are projected onto a sphere, whose center and radius are set to minimize the mean square residual of network points from the surface of that sphere.
A convex hull of points is then constructed from these projected positions on the sphere, using the convhulln routine in Matlab, yielding a set of triangles. Triangles are rejected if their center is more than $0.3\mu\text{m}$ from the sphere surface or the orientation of their normal vector is more than $40^\circ$ from the radial direction.
This procedure effectively removes triangles spanning across large sphere regions not covered by the mitochondrial network.
The areas of the remaining triangles are summed and used as an effective area spanned by the mitochondrial network.

For the ER structures, a convex hull is found from the two-dimensional points along the largest connected component of each network. The total area of the convex hull is then used for the effective area of the endoplasmic reticulum network. 

\subsection{Approximating search times with percolation theory}

We consider a fully connected $n\times n$ square lattice of network nodes on a unit square, giving $N=n^2$ total nodes connected to nearest neighbors by edges of length $\ell=1/(n-1)$. The complete lattice has $E_{\text{max}}=2n(n-1)$ total edges. Edges are removed from the network until a number $\Gamma$ of loops remain, without disconnecting any nodes. The number of edges remaining is $E = n^2-1 + \Gamma$. The fraction of edges that remain is $p = E/E_{\text{max}}$,
\begin{equation}
p = \frac{1}{2} + \frac{1}{2n} + \frac{\Gamma}{2n(n-1)} \ .
\end{equation}
The critical bond probability for percolation on a square lattice is $p_{\text{c}} = 1/2$~\cite{stauffer94}, such that our lattice has $p>p_{\text{c}}$. For $p>p_{\text{c}}$ but remaining near $p_{\text{c}}$, the diffusivity depends on the bond probability as $D\sim(p-p_{\text{c}})^{\mu}$, where $\mu\simeq 1.30$~\cite{stauffer94}. The diffusivity on the network is
\begin{subequations}
\begin{align}
D \sim &\left(\frac{1}{2} + \frac{1}{2n} + \frac{\Gamma}{2n(n-1)} - p_{\text{c}}\right)^{\mu}\\
\sim &\left(\frac{1}{2n} + \frac{\Gamma}{2n(n-1)}\right)^{\mu} \label{eq:diffusivity} \ .
\end{align}
\end{subequations}
Note that for a fixed loop number $\Gamma$, increasing the total edge length corresponds to increasing the lattice density $n$, and hence decreasing the effective diffusivity. Specifically, the total edge length for the network is given by
 $L = \ell E = n + 1 + \Gamma/(n-1)$. We can then express the lattice density in terms of our control parameters $L$ and $\Gamma$ according to,
\begin{equation}
\label{eq:n}
n = \frac{L}{2}\left(1 + \sqrt{1 - \frac{4(L+\Gamma-1)}{L^2}}\right) \ .
\end{equation}

Assuming a dense lattice system with $L\gg 1$ and $\Gamma \ll L^2$, we get the scaling $n\sim L$. This can be plugged into the diffusivity (Eq.~\ref{eq:diffusivity}) to show
\begin{subequations}
	\begin{align}
		D \sim &\left(\frac{L + \Gamma}{L^2}\right)^{\mu} \label{eq:diffusivity2}
	\end{align}
\end{subequations}
For a highly disconnected system with low loop number ($\Gamma \ll L$), the diffusivity scales as $D \sim \Gamma^0L^{-\mu}$. In the opposite extreme of high loop number ($L\ll \Gamma \ll L^2$), the diffusivity scales as $D \sim \Gamma^\mu L^{-2\mu}$.


%

Random walks on the largest connected component of a planar lattice above the percolation transition are expected to show the universal scaling behavior associated with diffusion in two dimensions, at sufficiently large length scales~\cite{stauffer94}. The target-site search time for a two-dimensional random walk with unit time steps is known to scale with the number of sites $T_\text{step}\sim N$, neglecting a logarithmic correction term~\cite{condamin08,benichou10}. For our system, the particles diffuse with diffusivity $D$ along edges of length $\ell$, so that the characteristic time to traverse each edge scales as $\ell^2/D$. Consequently, the overall expected search time is
\begin{equation}
\label{eq:inversediffusion}
T \sim \frac{\ell^2}{D} N \sim \frac{(n-1)^{-2}}{D} n^2 \sim \frac{1}{D} \ .
\end{equation}

We thus expect the search time to vary as $T\sim L^{\mu}$ for nearly loop-less networks, and $T\sim \Gamma^{-\mu}L^{2\mu}$ for networks with high loop numbers.

A similar argument can be used to relate the first passage time $T$ and the effective diffusivity $D$ for a compact search process where the underlying particle dynamics is subdiffusive, with $\left<x^2\right>\sim t^\alpha$ on each edge. Namely, if the dimensionality of the walk ($d_{\text{w}}=2/\alpha$) is less than the dimensionality of the domain ($d_{\text{f}}$) then we have $T_{\text{step}}\sim N^{d_{\text{w}}/d_{\text{f}}}$~\cite{benichou10}. For a planar network above the percolation transition, $d_{\text{f}} = 2$ and Eq.~\ref{eq:inversediffusion} can be modified to
\begin{equation}
T \sim \frac{\ell^{2/\alpha}}{D}N^{\alpha/2} \sim \frac{(n-1)^{-2/\alpha}}{D}\left(n^2\right)^{\frac{2/\alpha}{2}} \sim \frac{1}{D}.
\end{equation}
The coefficient $D$ in this case characterizes the large-scale spreading of particles over the network structure. The dependence of $D$ on network connectivity in a cluster near percolation ({\em i.e.:} the scaling exponent $\mu$) is likely to be altered for subdiffusive motion. However, we do not address this behavior here, focusing instead on diffusive search processes that are expected to be relevant for a variety of proteins in the ER and mitochondrial networks.


\subsection{Search times on simple loop-less network topologies}


Self-similar, hierarchically branched tree networks are constructed iteratively by attaching $m$ additional branches to the center of each branch in an existing tree. The number of steps $S$ to find a central target on a tree generated by $g$ iterations scales as 
\begin{equation}
S_g\sim N_g^{1 + \log2/\log(m+2)},
\end{equation}
where $N_g =  (m+2)^g + 1$ is the number of nodes in the tree~\cite{lin2010determining}. The number of tree edges is given by $K_g = (m+2)^g$, so $K_g\sim N_g$. 

If such a hierarchical tree network is constrained to a domain of unit radius, the edge lengths of the tree must become shorter with each iteration, scaling as $\ell_g = 2^{-g}$. The total edge length will then be $L_g = K_g\ell_g \sim [(m+2)/2]^g$.
The time required to diffuse across each edge is $\Delta t_g\sim\ell_g^2$. Overall, the total time for diffusive search to the target will scale as 
\begin{equation}
\begin{split}
T_g = \Delta t_g S_g & \sim \ell^{1 -\frac{\log 2}{\log(m+2)} } L_g^{1 + \frac{\log 2}{\log(m+2)}} \\
& \sim L_g \left[ 2^ {-g + \frac{g\log 2}{\log (m+2)}} \right(\frac{m+2}{2}\left)^{\frac{g\log 2}{\log (m+2)}}\right] \\
& \sim L_g
\end{split}
\end{equation}

Consequently, the time for diffusive search over a fractal tree network scaled to fit within a domain of fixed spatial extent should scale as $T\sim L$, as indicated in the main text.

\section{Additional information}

{\bf Competing interests}: The authors declare no competing interests.

{\bf Author contributions}: A.I.B.\ and E.F.K.\ designed the study, developed the model, performed the calculations, and revised the manuscript. A.I.B.\ analyzed the data and wrote the initial manuscript draft. L.M.W.\ cultured and imaged cells. All authors discussed the results and commented on the manuscript.

{\bf Data availability}: 
Datasets generated and analysed during the current study are available from the corresponding author upon request. Software for computing mean first passage times is available in a GitHub repository at: \url{https://github.com/lenafabr/networkMFPT}.


\begin{acknowledgments}
{\bf Acknowledgments:} This work was supported by the Alfred P. Sloan Foundation (EFK), the Hellman Fellows Fund (EFK) and the National Institutes of Health (F32GM116371 to LMW and GM083977 to Gia Voeltz). The authors thank Gia Voeltz (University of Colorado, Boulder) for facilities and Gia Voeltz, Matheus Viana, Susanne Rafelski (Allen Institute), Saurabh Mogre, and Anamika Agrawal (UCSD Physics) for useful discussion and feedback.
\end{acknowledgments}


\begin{thebibliography}{83}%
	\makeatletter
	\providecommand \@ifxundefined [1]{%
		\@ifx{#1\undefined}
	}%
	\providecommand \@ifnum [1]{%
		\ifnum #1\expandafter \@firstoftwo
		\else \expandafter \@secondoftwo
		\fi
	}%
	\providecommand \@ifx [1]{%
		\ifx #1\expandafter \@firstoftwo
		\else \expandafter \@secondoftwo
		\fi
	}%
	\providecommand \natexlab [1]{#1}%
	\providecommand \enquote  [1]{``#1''}%
	\providecommand \bibnamefont  [1]{#1}%
	\providecommand \bibfnamefont [1]{#1}%
	\providecommand \citenamefont [1]{#1}%
	\providecommand \href@noop [0]{\@secondoftwo}%
	\providecommand \href [0]{\begingroup \@sanitize@url \@href}%
	\providecommand \@href[1]{\@@startlink{#1}\@@href}%
	\providecommand \@@href[1]{\endgroup#1\@@endlink}%
	\providecommand \@sanitize@url [0]{\catcode `\\12\catcode `\$12\catcode
		`\&12\catcode `\#12\catcode `\^12\catcode `\_12\catcode `\%12\relax}%
	\providecommand \@@startlink[1]{}%
	\providecommand \@@endlink[0]{}%
	\providecommand \url  [0]{\begingroup\@sanitize@url \@url }%
	\providecommand \@url [1]{\endgroup\@href {#1}{\urlprefix }}%
	\providecommand \urlprefix  [0]{URL }%
	\providecommand \Eprint [0]{\href }%
	\providecommand \doibase [0]{http://dx.doi.org/}%
	\providecommand \selectlanguage [0]{\@gobble}%
	\providecommand \bibinfo  [0]{\@secondoftwo}%
	\providecommand \bibfield  [0]{\@secondoftwo}%
	\providecommand \translation [1]{[#1]}%
	\providecommand \BibitemOpen [0]{}%
	\providecommand \bibitemStop [0]{}%
	\providecommand \bibitemNoStop [0]{.\EOS\space}%
	\providecommand \EOS [0]{\spacefactor3000\relax}%
	\providecommand \BibitemShut  [1]{\csname bibitem#1\endcsname}%
	\let\auto@bib@innerbib\@empty
	\bibitem [{\citenamefont {Boccaletti}\ \emph {et~al.}(2006)\citenamefont
		{Boccaletti}, \citenamefont {Latora}, \citenamefont {Moreno}, \citenamefont
		{Chavez},\ and\ \citenamefont {Hwang}}]{boccaletti06}%
	\BibitemOpen
	\bibfield  {author} {\bibinfo {author} {\bibfnamefont {S.}~\bibnamefont
			{Boccaletti}}, \bibinfo {author} {\bibfnamefont {V.}~\bibnamefont {Latora}},
		\bibinfo {author} {\bibfnamefont {Y.}~\bibnamefont {Moreno}}, \bibinfo
		{author} {\bibfnamefont {M.}~\bibnamefont {Chavez}}, \ and\ \bibinfo {author}
		{\bibfnamefont {D.-U.}\ \bibnamefont {Hwang}},\ }\href@noop {} {\bibfield
		{journal} {\bibinfo  {journal} {Phys.\ Rep.}\ }\textbf {\bibinfo {volume}
			{424}},\ \bibinfo {pages} {175} (\bibinfo {year} {2006})}\BibitemShut
	{NoStop}%
	\bibitem [{\citenamefont {Durand}(2007)}]{durand07}%
	\BibitemOpen
	\bibfield  {author} {\bibinfo {author} {\bibfnamefont {M.}~\bibnamefont
			{Durand}},\ }\href@noop {} {\bibfield  {journal} {\bibinfo  {journal} {Phys.\
				Rev.\ Lett.}\ }\textbf {\bibinfo {volume} {98}},\ \bibinfo {pages} {088701}
		(\bibinfo {year} {2007})}\BibitemShut {NoStop}%
	\bibitem [{\citenamefont {Cheianov}\ \emph {et~al.}(2007)\citenamefont
		{Cheianov}, \citenamefont {Fal’ko}, \citenamefont {Altshuler},\ and\
		\citenamefont {Aleiner}}]{cheianov07}%
	\BibitemOpen
	\bibfield  {author} {\bibinfo {author} {\bibfnamefont {V.~V.}\ \bibnamefont
			{Cheianov}}, \bibinfo {author} {\bibfnamefont {V.~I.}\ \bibnamefont
			{Fal’ko}}, \bibinfo {author} {\bibfnamefont {B.~L.}\ \bibnamefont
			{Altshuler}}, \ and\ \bibinfo {author} {\bibfnamefont {I.~L.}\ \bibnamefont
			{Aleiner}},\ }\href@noop {} {\bibfield  {journal} {\bibinfo  {journal}
			{Phys.\ Rev.\ Lett.}\ }\textbf {\bibinfo {volume} {99}},\ \bibinfo {pages}
		{176801} (\bibinfo {year} {2007})}\BibitemShut {NoStop}%
	\bibitem [{\citenamefont {Bashan}\ \emph {et~al.}(2012)\citenamefont {Bashan},
		\citenamefont {Bartsch}, \citenamefont {Kantelhardt}, \citenamefont
		{Havlin},\ and\ \citenamefont {Ivanov}}]{bashan2012network}%
	\BibitemOpen
	\bibfield  {author} {\bibinfo {author} {\bibfnamefont {A.}~\bibnamefont
			{Bashan}}, \bibinfo {author} {\bibfnamefont {R.~P.}\ \bibnamefont {Bartsch}},
		\bibinfo {author} {\bibfnamefont {J.~W.}\ \bibnamefont {Kantelhardt}},
		\bibinfo {author} {\bibfnamefont {S.}~\bibnamefont {Havlin}}, \ and\ \bibinfo
		{author} {\bibfnamefont {P.~C.}\ \bibnamefont {Ivanov}},\ }\href@noop {}
	{\bibfield  {journal} {\bibinfo  {journal} {Nature communications}\ }\textbf
		{\bibinfo {volume} {3}},\ \bibinfo {pages} {1} (\bibinfo {year}
		{2012})}\BibitemShut {NoStop}%
	\bibitem [{\citenamefont {Ivanov}\ \emph {et~al.}(2016)\citenamefont {Ivanov},
		\citenamefont {Liu},\ and\ \citenamefont {Bartsch}}]{ivanov2016focus}%
	\BibitemOpen
	\bibfield  {author} {\bibinfo {author} {\bibfnamefont {P.~C.}\ \bibnamefont
			{Ivanov}}, \bibinfo {author} {\bibfnamefont {K.~K.}\ \bibnamefont {Liu}}, \
		and\ \bibinfo {author} {\bibfnamefont {R.~P.}\ \bibnamefont {Bartsch}},\
	}\href@noop {} {\bibfield  {journal} {\bibinfo  {journal} {New journal of
			physics}\ }\textbf {\bibinfo {volume} {18}},\ \bibinfo {pages} {100201}
	(\bibinfo {year} {2016})}\BibitemShut {NoStop}%
\bibitem [{\citenamefont {Wasserman}\ and\ \citenamefont
	{Faust}(1994)}]{wasserman94}%
\BibitemOpen
\bibfield  {author} {\bibinfo {author} {\bibfnamefont {S.}~\bibnamefont
		{Wasserman}}\ and\ \bibinfo {author} {\bibfnamefont {K.}~\bibnamefont
		{Faust}},\ }\href@noop {} {\emph {\bibinfo {title} {{Social network analysis:
				Methods and applications}}}}\ (\bibinfo  {publisher} {Cambridge University
	Press},\ \bibinfo {year} {1994})\BibitemShut {NoStop}%
\bibitem [{\citenamefont {Farrell}\ \emph {et~al.}(2018)\citenamefont
	{Farrell}, \citenamefont {Mitnitski}, \citenamefont {Theou}, \citenamefont
	{Rockwood},\ and\ \citenamefont {Rutenberg}}]{farrell18}%
\BibitemOpen
\bibfield  {author} {\bibinfo {author} {\bibfnamefont {S.~G.}\ \bibnamefont
		{Farrell}}, \bibinfo {author} {\bibfnamefont {A.~B.}\ \bibnamefont
		{Mitnitski}}, \bibinfo {author} {\bibfnamefont {O.}~\bibnamefont {Theou}},
	\bibinfo {author} {\bibfnamefont {K.}~\bibnamefont {Rockwood}}, \ and\
	\bibinfo {author} {\bibfnamefont {A.~D.}\ \bibnamefont {Rutenberg}},\
}\href@noop {} {\bibfield  {journal} {\bibinfo  {journal} {Phys.\ Rev.\ E}\
}\textbf {\bibinfo {volume} {98}},\ \bibinfo {pages} {032302} (\bibinfo
{year} {2018})}\BibitemShut {NoStop}%
\bibitem [{\citenamefont {Masuda}\ \emph {et~al.}(2017)\citenamefont {Masuda},
	\citenamefont {Porter},\ and\ \citenamefont {Lambiotte}}]{masuda17}%
\BibitemOpen
\bibfield  {author} {\bibinfo {author} {\bibfnamefont {N.}~\bibnamefont
		{Masuda}}, \bibinfo {author} {\bibfnamefont {M.}~\bibnamefont {Porter}}, \
	and\ \bibinfo {author} {\bibfnamefont {R.}~\bibnamefont {Lambiotte}},\
}\href@noop {} {\bibfield  {journal} {\bibinfo  {journal} {Phys.\ Rep.}\
}\textbf {\bibinfo {volume} {716-717}},\ \bibinfo {pages} {1} (\bibinfo
{year} {2017})}\BibitemShut {NoStop}%
\bibitem [{\citenamefont {Pu}\ \emph {et~al.}(2015)\citenamefont {Pu},
	\citenamefont {Li},\ and\ \citenamefont {Yang}}]{pu15}%
\BibitemOpen
\bibfield  {author} {\bibinfo {author} {\bibfnamefont {C.}~\bibnamefont
		{Pu}}, \bibinfo {author} {\bibfnamefont {S.}~\bibnamefont {Li}}, \ and\
	\bibinfo {author} {\bibfnamefont {J.}~\bibnamefont {Yang}},\ }\href@noop {}
{\bibfield  {journal} {\bibinfo  {journal} {Physica A}\ }\textbf {\bibinfo
		{volume} {432}},\ \bibinfo {pages} {230} (\bibinfo {year}
	{2015})}\BibitemShut {NoStop}%
\bibitem [{\citenamefont {Perna}\ and\ \citenamefont {Latty}(2014)}]{perna14}%
\BibitemOpen
\bibfield  {author} {\bibinfo {author} {\bibfnamefont {A.}~\bibnamefont
		{Perna}}\ and\ \bibinfo {author} {\bibfnamefont {T.}~\bibnamefont {Latty}},\
}\href@noop {} {\bibfield  {journal} {\bibinfo  {journal} {J.\ R.\ Soc.\
		Interface.}\ }\textbf {\bibinfo {volume} {11}},\ \bibinfo {pages} {20140334}
(\bibinfo {year} {2014})}\BibitemShut {NoStop}%
\bibitem [{\citenamefont {Chavez}\ \emph {et~al.}(2010)\citenamefont {Chavez},
	\citenamefont {Valencia}, \citenamefont {Navarro}, \citenamefont {Latora},\
	and\ \citenamefont {Martinerie}}]{chavez10}%
\BibitemOpen
\bibfield  {author} {\bibinfo {author} {\bibfnamefont {M.}~\bibnamefont
		{Chavez}}, \bibinfo {author} {\bibfnamefont {M.}~\bibnamefont {Valencia}},
	\bibinfo {author} {\bibfnamefont {V.}~\bibnamefont {Navarro}}, \bibinfo
	{author} {\bibfnamefont {V.}~\bibnamefont {Latora}}, \ and\ \bibinfo {author}
	{\bibfnamefont {J.}~\bibnamefont {Martinerie}},\ }\href@noop {} {\bibfield
	{journal} {\bibinfo  {journal} {Phys.\ Rev.\ Lett.}\ }\textbf {\bibinfo
		{volume} {104}},\ \bibinfo {pages} {118701} (\bibinfo {year}
	{2010})}\BibitemShut {NoStop}%
\bibitem [{\citenamefont {Nelson}(1999)}]{nelson99}%
\BibitemOpen
\bibfield  {author} {\bibinfo {author} {\bibfnamefont {J.}~\bibnamefont
		{Nelson}},\ }\href@noop {} {\bibfield  {journal} {\bibinfo  {journal} {Phys.\
			Rev.\ B}\ }\textbf {\bibinfo {volume} {59}},\ \bibinfo {pages} {15374}
	(\bibinfo {year} {1999})}\BibitemShut {NoStop}%
\bibitem [{\citenamefont {Burioni}\ and\ \citenamefont
	{Cassi}(2005)}]{burioni05}%
\BibitemOpen
\bibfield  {author} {\bibinfo {author} {\bibfnamefont {R.}~\bibnamefont
		{Burioni}}\ and\ \bibinfo {author} {\bibfnamefont {D.}~\bibnamefont
		{Cassi}},\ }\href@noop {} {\bibfield  {journal} {\bibinfo  {journal} {J.
			Phys. A: Math. Gen.}\ }\textbf {\bibinfo {volume} {38}},\ \bibinfo {pages}
	{R45} (\bibinfo {year} {2005})}\BibitemShut {NoStop}%
\bibitem [{\citenamefont {Hwang}\ \emph {et~al.}(2012)\citenamefont {Hwang},
	\citenamefont {Lee},\ and\ \citenamefont {Kahng}}]{hwang12}%
\BibitemOpen
\bibfield  {author} {\bibinfo {author} {\bibfnamefont {S.}~\bibnamefont
		{Hwang}}, \bibinfo {author} {\bibfnamefont {D.-S.}\ \bibnamefont {Lee}}, \
	and\ \bibinfo {author} {\bibfnamefont {B.}~\bibnamefont {Kahng}},\
}\href@noop {} {\bibfield  {journal} {\bibinfo  {journal} {Phys.\ Rev.\
		Lett.}\ }\textbf {\bibinfo {volume} {109}},\ \bibinfo {pages} {088701}
(\bibinfo {year} {2012})}\BibitemShut {NoStop}%
\bibitem [{\citenamefont {Noh}\ and\ \citenamefont {Rieger}(2004)}]{noh04}%
\BibitemOpen
\bibfield  {author} {\bibinfo {author} {\bibfnamefont {J.~D.}\ \bibnamefont
		{Noh}}\ and\ \bibinfo {author} {\bibfnamefont {H.}~\bibnamefont {Rieger}},\
}\href@noop {} {\bibfield  {journal} {\bibinfo  {journal} {Phys.\ Rev.\
		Lett.}\ }\textbf {\bibinfo {volume} {92}},\ \bibinfo {pages} {118701}
(\bibinfo {year} {2004})}\BibitemShut {NoStop}%
\bibitem [{\citenamefont {Barthelemy}(2011)}]{barthelemy11}%
\BibitemOpen
\bibfield  {author} {\bibinfo {author} {\bibfnamefont {M.}~\bibnamefont
		{Barthelemy}},\ }\href@noop {} {\bibfield  {journal} {\bibinfo  {journal}
		{Phys.\ Rep.}\ }\textbf {\bibinfo {volume} {499}},\ \bibinfo {pages} {1}
	(\bibinfo {year} {2011})}\BibitemShut {NoStop}%
\bibitem [{\citenamefont {Redner}(2001)}]{redner01}%
\BibitemOpen
\bibfield  {author} {\bibinfo {author} {\bibfnamefont {S.}~\bibnamefont
		{Redner}},\ }\href@noop {} {\emph {\bibinfo {title} {{A guide to
				first-passage processes}}}}\ (\bibinfo  {publisher} {Cambridge University
	Press},\ \bibinfo {year} {2001})\BibitemShut {NoStop}%
\bibitem [{\citenamefont {Ben-Avraham}\ and\ \citenamefont
	{Havlin}(2000)}]{ben2000diffusion}%
\BibitemOpen
\bibfield  {author} {\bibinfo {author} {\bibfnamefont {D.}~\bibnamefont
		{Ben-Avraham}}\ and\ \bibinfo {author} {\bibfnamefont {S.}~\bibnamefont
		{Havlin}},\ }\href@noop {} {\emph {\bibinfo {title} {Diffusion and reactions
			in fractals and disordered systems}}}\ (\bibinfo  {publisher} {Cambridge
	university press},\ \bibinfo {year} {2000})\BibitemShut {NoStop}%
\bibitem [{\citenamefont {Benichou}\ \emph {et~al.}(2010)\citenamefont
	{Benichou}, \citenamefont {Chevalier}, \citenamefont {Klafter}, \citenamefont
	{B.Meyer},\ and\ \citenamefont {Voituriez}}]{benichou10}%
\BibitemOpen
\bibfield  {author} {\bibinfo {author} {\bibfnamefont {O.}~\bibnamefont
		{Benichou}}, \bibinfo {author} {\bibfnamefont {C.}~\bibnamefont {Chevalier}},
	\bibinfo {author} {\bibfnamefont {J.}~\bibnamefont {Klafter}}, \bibinfo
	{author} {\bibnamefont {B.Meyer}}, \ and\ \bibinfo {author} {\bibfnamefont
		{R.}~\bibnamefont {Voituriez}},\ }\href@noop {} {\bibfield  {journal}
	{\bibinfo  {journal} {Nat.\ Chem.}\ }\textbf {\bibinfo {volume} {2}},\
	\bibinfo {pages} {472} (\bibinfo {year} {2010})}\BibitemShut {NoStop}%
\bibitem [{\citenamefont {Shen}\ and\ \citenamefont
	{Chen}(2007)}]{shen2007critical}%
\BibitemOpen
\bibfield  {author} {\bibinfo {author} {\bibfnamefont {L.}~\bibnamefont
		{Shen}}\ and\ \bibinfo {author} {\bibfnamefont {Z.}~\bibnamefont {Chen}},\
}\href@noop {} {\bibfield  {journal} {\bibinfo  {journal} {Chemical
		Engineering Science}\ }\textbf {\bibinfo {volume} {62}},\ \bibinfo {pages}
{3748} (\bibinfo {year} {2007})}\BibitemShut {NoStop}%
\bibitem [{\citenamefont {Havlin}\ and\ \citenamefont
	{Ben-Avraham}(1987)}]{havlin87}%
\BibitemOpen
\bibfield  {author} {\bibinfo {author} {\bibfnamefont {S.}~\bibnamefont
		{Havlin}}\ and\ \bibinfo {author} {\bibfnamefont {D.}~\bibnamefont
		{Ben-Avraham}},\ }\href@noop {} {\bibfield  {journal} {\bibinfo  {journal}
		{Adv.\ Phys.}\ }\textbf {\bibinfo {volume} {36}},\ \bibinfo {pages} {695}
	(\bibinfo {year} {1987})}\BibitemShut {NoStop}%
\bibitem [{\citenamefont {Condamin}\ \emph {et~al.}(2007)\citenamefont
	{Condamin}, \citenamefont {Benichou}, \citenamefont {Tejedor}, \citenamefont
	{Voituriez},\ and\ \citenamefont {Klafter}}]{condamin07}%
\BibitemOpen
\bibfield  {author} {\bibinfo {author} {\bibfnamefont {S.}~\bibnamefont
		{Condamin}}, \bibinfo {author} {\bibfnamefont {O.}~\bibnamefont {Benichou}},
	\bibinfo {author} {\bibfnamefont {V.}~\bibnamefont {Tejedor}}, \bibinfo
	{author} {\bibfnamefont {R.}~\bibnamefont {Voituriez}}, \ and\ \bibinfo
	{author} {\bibfnamefont {J.}~\bibnamefont {Klafter}},\ }\href@noop {}
{\bibfield  {journal} {\bibinfo  {journal} {Nature}\ }\textbf {\bibinfo
		{volume} {450}},\ \bibinfo {pages} {77} (\bibinfo {year} {2007})}\BibitemShut
{NoStop}%
\bibitem [{\citenamefont {Benichou}\ and\ \citenamefont
	{Voituriez}(2014)}]{benichou14}%
\BibitemOpen
\bibfield  {author} {\bibinfo {author} {\bibfnamefont {O.}~\bibnamefont
		{Benichou}}\ and\ \bibinfo {author} {\bibfnamefont {R.}~\bibnamefont
		{Voituriez}},\ }\href@noop {} {\bibfield  {journal} {\bibinfo  {journal}
		{Phys.\ Rep.}\ }\textbf {\bibinfo {volume} {539}},\ \bibinfo {pages} {225}
	(\bibinfo {year} {2014})}\BibitemShut {NoStop}%
\bibitem [{\citenamefont {Chupeau}\ \emph {et~al.}(2015)\citenamefont
	{Chupeau}, \citenamefont {Benichou},\ and\ \citenamefont
	{Voituriez}}]{chupeau15}%
\BibitemOpen
\bibfield  {author} {\bibinfo {author} {\bibfnamefont {M.}~\bibnamefont
		{Chupeau}}, \bibinfo {author} {\bibfnamefont {O.}~\bibnamefont {Benichou}}, \
	and\ \bibinfo {author} {\bibfnamefont {R.}~\bibnamefont {Voituriez}},\
}\href@noop {} {\bibfield  {journal} {\bibinfo  {journal} {Nat.\ Phys.}\
}\textbf {\bibinfo {volume} {11}},\ \bibinfo {pages} {844} (\bibinfo {year}
{2015})}\BibitemShut {NoStop}%
\bibitem [{\citenamefont {Westrate}\ \emph {et~al.}(2015)\citenamefont
	{Westrate}, \citenamefont {Lee}, \citenamefont {Prinz},\ and\ \citenamefont
	{Voeltz}}]{westrate15}%
\BibitemOpen
\bibfield  {author} {\bibinfo {author} {\bibfnamefont {L.~M.}\ \bibnamefont
		{Westrate}}, \bibinfo {author} {\bibfnamefont {J.~E.}\ \bibnamefont {Lee}},
	\bibinfo {author} {\bibfnamefont {W.~A.}\ \bibnamefont {Prinz}}, \ and\
	\bibinfo {author} {\bibfnamefont {G.~K.}\ \bibnamefont {Voeltz}},\
}\href@noop {} {\bibfield  {journal} {\bibinfo  {journal} {Annu.\ Rev.\
		Biochem.}\ }\textbf {\bibinfo {volume} {84}},\ \bibinfo {pages} {791}
(\bibinfo {year} {2015})}\BibitemShut {NoStop}%
\bibitem [{\citenamefont {Schwarz}\ and\ \citenamefont
	{Blower}(2016)}]{schwarz16}%
\BibitemOpen
\bibfield  {author} {\bibinfo {author} {\bibfnamefont {D.~S.}\ \bibnamefont
		{Schwarz}}\ and\ \bibinfo {author} {\bibfnamefont {M.~D.}\ \bibnamefont
		{Blower}},\ }\href@noop {} {\bibfield  {journal} {\bibinfo  {journal} {Cell.\
			Mol.\ Life Sci.}\ }\textbf {\bibinfo {volume} {73}},\ \bibinfo {pages} {79}
	(\bibinfo {year} {2016})}\BibitemShut {NoStop}%
\bibitem [{\citenamefont {Collins}\ \emph {et~al.}(2002)\citenamefont
	{Collins}, \citenamefont {Berridge}, \citenamefont {Lipp},\ and\
	\citenamefont {Bootman}}]{collins02}%
\BibitemOpen
\bibfield  {author} {\bibinfo {author} {\bibfnamefont {T.~J.}\ \bibnamefont
		{Collins}}, \bibinfo {author} {\bibfnamefont {M.~J.}\ \bibnamefont
		{Berridge}}, \bibinfo {author} {\bibfnamefont {P.}~\bibnamefont {Lipp}}, \
	and\ \bibinfo {author} {\bibfnamefont {M.~D.}\ \bibnamefont {Bootman}},\
}\href@noop {} {\bibfield  {journal} {\bibinfo  {journal} {EMBO J.}\ }\textbf
{\bibinfo {volume} {21}},\ \bibinfo {pages} {1616} (\bibinfo {year}
{2002})}\BibitemShut {NoStop}%
\bibitem [{\citenamefont {Rafelski}\ \emph {et~al.}(2012)\citenamefont
	{Rafelski}, \citenamefont {Viana}, \citenamefont {Zhang}, \citenamefont
	{Chan}, \citenamefont {Thorn}, \citenamefont {Yam}, \citenamefont {Fung},
	\citenamefont {Li}, \citenamefont {da~F.~Costa},\ and\ \citenamefont
	{Marshall}}]{rafelski12}%
\BibitemOpen
\bibfield  {author} {\bibinfo {author} {\bibfnamefont {S.~M.}\ \bibnamefont
		{Rafelski}}, \bibinfo {author} {\bibfnamefont {M.~P.}\ \bibnamefont {Viana}},
	\bibinfo {author} {\bibfnamefont {Y.}~\bibnamefont {Zhang}}, \bibinfo
	{author} {\bibfnamefont {Y.-H.~M.}\ \bibnamefont {Chan}}, \bibinfo {author}
	{\bibfnamefont {K.~S.}\ \bibnamefont {Thorn}}, \bibinfo {author}
	{\bibfnamefont {P.}~\bibnamefont {Yam}}, \bibinfo {author} {\bibfnamefont
		{J.~C.}\ \bibnamefont {Fung}}, \bibinfo {author} {\bibfnamefont
		{H.}~\bibnamefont {Li}}, \bibinfo {author} {\bibfnamefont {L.}~\bibnamefont
		{da~F.~Costa}}, \ and\ \bibinfo {author} {\bibfnamefont {W.~F.}\ \bibnamefont
		{Marshall}},\ }\href@noop {} {\bibfield  {journal} {\bibinfo  {journal}
		{Science}\ }\textbf {\bibinfo {volume} {338}},\ \bibinfo {pages} {822}
	(\bibinfo {year} {2012})}\BibitemShut {NoStop}%
\bibitem [{\citenamefont {Speckner}\ \emph {et~al.}(2018)\citenamefont
	{Speckner}, \citenamefont {Stadler},\ and\ \citenamefont
	{Weiss}}]{speckner18}%
\BibitemOpen
\bibfield  {author} {\bibinfo {author} {\bibfnamefont {K.}~\bibnamefont
		{Speckner}}, \bibinfo {author} {\bibfnamefont {L.}~\bibnamefont {Stadler}}, \
	and\ \bibinfo {author} {\bibfnamefont {M.}~\bibnamefont {Weiss}},\
}\href@noop {} {\bibfield  {journal} {\bibinfo  {journal} {Phys.\ Rev.\ E}\
}\textbf {\bibinfo {volume} {98}},\ \bibinfo {pages} {012406} (\bibinfo
{year} {2018})}\BibitemShut {NoStop}%
\bibitem [{\citenamefont {Harwig}\ \emph {et~al.}(2018)\citenamefont {Harwig},
	\citenamefont {Viana}, \citenamefont {Egner}, \citenamefont {Harwig},
	\citenamefont {Widlansky}, \citenamefont {Rafelski},\ and\ \citenamefont
	{Hill}}]{harwig18}%
\BibitemOpen
\bibfield  {author} {\bibinfo {author} {\bibfnamefont {M.~C.}\ \bibnamefont
		{Harwig}}, \bibinfo {author} {\bibfnamefont {M.~P.}\ \bibnamefont {Viana}},
	\bibinfo {author} {\bibfnamefont {J.~M.}\ \bibnamefont {Egner}}, \bibinfo
	{author} {\bibfnamefont {J.~J.}\ \bibnamefont {Harwig}}, \bibinfo {author}
	{\bibfnamefont {M.~E.}\ \bibnamefont {Widlansky}}, \bibinfo {author}
	{\bibfnamefont {S.~M.}\ \bibnamefont {Rafelski}}, \ and\ \bibinfo {author}
	{\bibfnamefont {R.~B.}\ \bibnamefont {Hill}},\ }\href@noop {} {\bibfield
	{journal} {\bibinfo  {journal} {Anal.\ Biochem.}\ }\textbf {\bibinfo {volume}
		{552}},\ \bibinfo {pages} {81} (\bibinfo {year} {2018})}\BibitemShut
{NoStop}%
\bibitem [{\citenamefont {Lin}\ \emph {et~al.}(2014)\citenamefont {Lin},
	\citenamefont {Zhang}, \citenamefont {Sparkes},\ and\ \citenamefont
	{Ashwin}}]{lin14}%
\BibitemOpen
\bibfield  {author} {\bibinfo {author} {\bibfnamefont {C.}~\bibnamefont
		{Lin}}, \bibinfo {author} {\bibfnamefont {Y.}~\bibnamefont {Zhang}}, \bibinfo
	{author} {\bibfnamefont {I.}~\bibnamefont {Sparkes}}, \ and\ \bibinfo
	{author} {\bibfnamefont {P.}~\bibnamefont {Ashwin}},\ }\href@noop {}
{\bibfield  {journal} {\bibinfo  {journal} {Biophys.\ J.}\ }\textbf {\bibinfo
		{volume} {107}},\ \bibinfo {pages} {763} (\bibinfo {year}
	{2014})}\BibitemShut {NoStop}%
\bibitem [{\citenamefont {Viana}\ \emph {et~al.}(ress)\citenamefont {Viana},
	\citenamefont {Brown}, \citenamefont {Mueller}, \citenamefont {Goul},
	\citenamefont {Koslover},\ and\ \citenamefont {Rafelski}}]{viana19}%
\BibitemOpen
\bibfield  {author} {\bibinfo {author} {\bibfnamefont {M.~P.}\ \bibnamefont
		{Viana}}, \bibinfo {author} {\bibfnamefont {A.~I.}\ \bibnamefont {Brown}},
	\bibinfo {author} {\bibfnamefont {I.~A.}\ \bibnamefont {Mueller}}, \bibinfo
	{author} {\bibfnamefont {C.}~\bibnamefont {Goul}}, \bibinfo {author}
	{\bibfnamefont {E.~F.}\ \bibnamefont {Koslover}}, \ and\ \bibinfo {author}
	{\bibfnamefont {S.~M.}\ \bibnamefont {Rafelski}},\ }\href@noop {} {\bibfield
	{journal} {\bibinfo  {journal} {Cell Syst.}\ } (\bibinfo {year} {In
		Press})}\BibitemShut {NoStop}%
\bibitem [{\citenamefont {Hughes}\ \emph {et~al.}(2009)\citenamefont {Hughes},
	\citenamefont {Budnik}, \citenamefont {Schmidt}, \citenamefont {Palmer},
	\citenamefont {Mantell}, \citenamefont {Noakes}, \citenamefont {Johnson},
	\citenamefont {Carter}, \citenamefont {Verkade}, \citenamefont {Watson},\
	and\ \citenamefont {Stephens}}]{hughes09}%
\BibitemOpen
\bibfield  {author} {\bibinfo {author} {\bibfnamefont {H.}~\bibnamefont
		{Hughes}}, \bibinfo {author} {\bibfnamefont {A.}~\bibnamefont {Budnik}},
	\bibinfo {author} {\bibfnamefont {K.}~\bibnamefont {Schmidt}}, \bibinfo
	{author} {\bibfnamefont {K.~J.}\ \bibnamefont {Palmer}}, \bibinfo {author}
	{\bibfnamefont {J.}~\bibnamefont {Mantell}}, \bibinfo {author} {\bibfnamefont
		{C.}~\bibnamefont {Noakes}}, \bibinfo {author} {\bibfnamefont
		{A.}~\bibnamefont {Johnson}}, \bibinfo {author} {\bibfnamefont {D.~A.}\
		\bibnamefont {Carter}}, \bibinfo {author} {\bibfnamefont {P.}~\bibnamefont
		{Verkade}}, \bibinfo {author} {\bibfnamefont {P.}~\bibnamefont {Watson}}, \
	and\ \bibinfo {author} {\bibfnamefont {D.~J.}\ \bibnamefont {Stephens}},\
}\href@noop {} {\bibfield  {journal} {\bibinfo  {journal} {J.\ Cell Sci.}\
}\textbf {\bibinfo {volume} {122}},\ \bibinfo {pages} {2924} (\bibinfo {year}
{2009})}\BibitemShut {NoStop}%
\bibitem [{\citenamefont {Ruhanen}\ \emph {et~al.}(2010)\citenamefont
	{Ruhanen}, \citenamefont {Borrie}, \citenamefont {Szabadkai}, \citenamefont
	{Tyynismaa}, \citenamefont {Jones}, \citenamefont {Kang}, \citenamefont
	{Taanman},\ and\ \citenamefont {Yasukawa}}]{ruhanen10}%
\BibitemOpen
\bibfield  {author} {\bibinfo {author} {\bibfnamefont {H.}~\bibnamefont
		{Ruhanen}}, \bibinfo {author} {\bibfnamefont {S.}~\bibnamefont {Borrie}},
	\bibinfo {author} {\bibfnamefont {G.}~\bibnamefont {Szabadkai}}, \bibinfo
	{author} {\bibfnamefont {H.}~\bibnamefont {Tyynismaa}}, \bibinfo {author}
	{\bibfnamefont {A.~W.~E.}\ \bibnamefont {Jones}}, \bibinfo {author}
	{\bibfnamefont {D.}~\bibnamefont {Kang}}, \bibinfo {author} {\bibfnamefont
		{J.-W.}\ \bibnamefont {Taanman}}, \ and\ \bibinfo {author} {\bibfnamefont
		{T.}~\bibnamefont {Yasukawa}},\ }\href@noop {} {\bibfield  {journal}
	{\bibinfo  {journal} {Biochim.\ Biophys.\ Acta}\ }\textbf {\bibinfo {volume}
		{1803}},\ \bibinfo {pages} {931} (\bibinfo {year} {2010})}\BibitemShut
{NoStop}%
\bibitem [{\citenamefont {English}\ and\ \citenamefont
	{Voeltz}(2013{\natexlab{a}})}]{english13}%
\BibitemOpen
\bibfield  {author} {\bibinfo {author} {\bibfnamefont {A.~R.}\ \bibnamefont
		{English}}\ and\ \bibinfo {author} {\bibfnamefont {G.~K.}\ \bibnamefont
		{Voeltz}},\ }\href@noop {} {\bibfield  {journal} {\bibinfo  {journal} {Cold
			Spring Harb.\ Perspect.\ Biol.}\ }\textbf {\bibinfo {volume} {5}},\ \bibinfo
	{pages} {a013227} (\bibinfo {year} {2013}{\natexlab{a}})}\BibitemShut
{NoStop}%
\bibitem [{\citenamefont {Rube}\ and\ \citenamefont {van~der
		Bliek}(2004)}]{rube04}%
\BibitemOpen
\bibfield  {author} {\bibinfo {author} {\bibfnamefont {D.~A.}\ \bibnamefont
		{Rube}}\ and\ \bibinfo {author} {\bibfnamefont {A.~M.}\ \bibnamefont {van~der
			Bliek}},\ }\href@noop {} {\bibfield  {journal} {\bibinfo  {journal} {Mol.\
			Cell.\ Biochem.}\ }\textbf {\bibinfo {volume} {256/257}},\ \bibinfo {pages}
	{331} (\bibinfo {year} {2004})}\BibitemShut {NoStop}%
\bibitem [{\citenamefont {Margineantu}\ \emph {et~al.}(2002)\citenamefont
	{Margineantu}, \citenamefont {Cox}, \citenamefont {Sundell}, \citenamefont
	{Sherwood}, \citenamefont {Beechem},\ and\ \citenamefont
	{Capaldi}}]{margineantu02}%
\BibitemOpen
\bibfield  {author} {\bibinfo {author} {\bibfnamefont {D.~H.}\ \bibnamefont
		{Margineantu}}, \bibinfo {author} {\bibfnamefont {W.~G.}\ \bibnamefont
		{Cox}}, \bibinfo {author} {\bibfnamefont {L.}~\bibnamefont {Sundell}},
	\bibinfo {author} {\bibfnamefont {S.~W.}\ \bibnamefont {Sherwood}}, \bibinfo
	{author} {\bibfnamefont {J.~M.}\ \bibnamefont {Beechem}}, \ and\ \bibinfo
	{author} {\bibfnamefont {R.~A.}\ \bibnamefont {Capaldi}},\ }\href@noop {}
{\bibfield  {journal} {\bibinfo  {journal} {Mitochondrion}\ }\textbf
	{\bibinfo {volume} {1}},\ \bibinfo {pages} {425–435} (\bibinfo {year}
	{2002})}\BibitemShut {NoStop}%
\bibitem [{\citenamefont {Shin}\ \emph {et~al.}(2016)\citenamefont {Shin},
	\citenamefont {Park}, \citenamefont {Kang}, \citenamefont {Wu}, \citenamefont
	{Choi},\ and\ \citenamefont {Shin}}]{shin16}%
\BibitemOpen
\bibfield  {author} {\bibinfo {author} {\bibfnamefont {J.~W.}\ \bibnamefont
		{Shin}}, \bibinfo {author} {\bibfnamefont {S.~H.}\ \bibnamefont {Park}},
	\bibinfo {author} {\bibfnamefont {Y.~G.}\ \bibnamefont {Kang}}, \bibinfo
	{author} {\bibfnamefont {Y.}~\bibnamefont {Wu}}, \bibinfo {author}
	{\bibfnamefont {H.~J.}\ \bibnamefont {Choi}}, \ and\ \bibinfo {author}
	{\bibfnamefont {J.-W.}\ \bibnamefont {Shin}},\ }\href@noop {} {\bibfield
	{journal} {\bibinfo  {journal} {PLoS One}\ }\textbf {\bibinfo {volume}
		{11}},\ \bibinfo {pages} {e0161015} (\bibinfo {year} {2016})}\BibitemShut
{NoStop}%
\bibitem [{\citenamefont {Willems}\ \emph {et~al.}(2009)\citenamefont
	{Willems}, \citenamefont {Smeitink},\ and\ \citenamefont
	{Koopman}}]{willems09}%
\BibitemOpen
\bibfield  {author} {\bibinfo {author} {\bibfnamefont {P.~H. G.~M.}\
		\bibnamefont {Willems}}, \bibinfo {author} {\bibfnamefont {J.~A.~M.}\
		\bibnamefont {Smeitink}}, \ and\ \bibinfo {author} {\bibfnamefont {W.~J.~H.}\
		\bibnamefont {Koopman}},\ }\href@noop {} {\bibfield  {journal} {\bibinfo
		{journal} {Int.\ J.\ Biochem.\ Cell Biol.}\ }\textbf {\bibinfo {volume}
		{41}},\ \bibinfo {pages} {1773} (\bibinfo {year} {2009})}\BibitemShut
{NoStop}%
\bibitem [{\citenamefont {Chiaradonna}\ \emph {et~al.}(2006)\citenamefont
	{Chiaradonna}, \citenamefont {Gaglio}, \citenamefont {Vanoni},\ and\
	\citenamefont {Alberghina}}]{chiaradonna06}%
\BibitemOpen
\bibfield  {author} {\bibinfo {author} {\bibfnamefont {F.}~\bibnamefont
		{Chiaradonna}}, \bibinfo {author} {\bibfnamefont {D.}~\bibnamefont {Gaglio}},
	\bibinfo {author} {\bibfnamefont {M.}~\bibnamefont {Vanoni}}, \ and\ \bibinfo
	{author} {\bibfnamefont {L.}~\bibnamefont {Alberghina}},\ }\href@noop {}
{\bibfield  {journal} {\bibinfo  {journal} {Biochim.\ Biophys.\ Acta}\
	}\textbf {\bibinfo {volume} {1757}},\ \bibinfo {pages} {1338} (\bibinfo
	{year} {2006})}\BibitemShut {NoStop}%
\bibitem [{\citenamefont {Ghosh}\ \emph {et~al.}(2018)\citenamefont {Ghosh},
	\citenamefont {Tran}, \citenamefont {Delbridge}, \citenamefont {Hickey},
	\citenamefont {Hanssen}, \citenamefont {Crampin},\ and\ \citenamefont
	{Rajagopal}}]{ghosh18}%
\BibitemOpen
\bibfield  {author} {\bibinfo {author} {\bibfnamefont {S.}~\bibnamefont
		{Ghosh}}, \bibinfo {author} {\bibfnamefont {K.}~\bibnamefont {Tran}},
	\bibinfo {author} {\bibfnamefont {L.~M.~D.}\ \bibnamefont {Delbridge}},
	\bibinfo {author} {\bibfnamefont {A.~J.~R.}\ \bibnamefont {Hickey}}, \bibinfo
	{author} {\bibfnamefont {E.}~\bibnamefont {Hanssen}}, \bibinfo {author}
	{\bibfnamefont {E.~J.}\ \bibnamefont {Crampin}}, \ and\ \bibinfo {author}
	{\bibfnamefont {V.}~\bibnamefont {Rajagopal}},\ }\href@noop {} {\bibfield
	{journal} {\bibinfo  {journal} {PLoS Comput.\ Biol.}\ }\textbf {\bibinfo
		{volume} {14}},\ \bibinfo {pages} {e1006640} (\bibinfo {year}
	{2018})}\BibitemShut {NoStop}%
\bibitem [{\citenamefont {Chen}\ \emph {et~al.}(2013)\citenamefont {Chen},
	\citenamefont {Novick},\ and\ \citenamefont {Ferro-Novick}}]{chen13}%
\BibitemOpen
\bibfield  {author} {\bibinfo {author} {\bibfnamefont {S.}~\bibnamefont
		{Chen}}, \bibinfo {author} {\bibfnamefont {P.}~\bibnamefont {Novick}}, \ and\
	\bibinfo {author} {\bibfnamefont {S.}~\bibnamefont {Ferro-Novick}},\
}\href@noop {} {\bibfield  {journal} {\bibinfo  {journal} {Curr.\ Opin.\ Cell
		Biol.}\ }\textbf {\bibinfo {volume} {25}},\ \bibinfo {pages} {428} (\bibinfo
{year} {2013})}\BibitemShut {NoStop}%
\bibitem [{\citenamefont {Zamponi}\ \emph {et~al.}(2018)\citenamefont
	{Zamponi}, \citenamefont {Zamponi}, \citenamefont {Cannas}, \citenamefont
	{Billoni}, \citenamefont {Helguera},\ and\ \citenamefont
	{Chialvo}}]{zamponi18}%
\BibitemOpen
\bibfield  {author} {\bibinfo {author} {\bibfnamefont {N.}~\bibnamefont
		{Zamponi}}, \bibinfo {author} {\bibfnamefont {E.}~\bibnamefont {Zamponi}},
	\bibinfo {author} {\bibfnamefont {S.~A.}\ \bibnamefont {Cannas}}, \bibinfo
	{author} {\bibfnamefont {O.~V.}\ \bibnamefont {Billoni}}, \bibinfo {author}
	{\bibfnamefont {P.~R.}\ \bibnamefont {Helguera}}, \ and\ \bibinfo {author}
	{\bibfnamefont {D.~R.}\ \bibnamefont {Chialvo}},\ }\href@noop {} {\bibfield
	{journal} {\bibinfo  {journal} {Sci.\ Rep.}\ }\textbf {\bibinfo {volume}
		{8}},\ \bibinfo {pages} {363} (\bibinfo {year} {2018})}\BibitemShut {NoStop}%
\bibitem [{\citenamefont {Dayel}\ \emph {et~al.}(1999)\citenamefont {Dayel},
	\citenamefont {Hom},\ and\ \citenamefont {Verkman}}]{dayel99}%
\BibitemOpen
\bibfield  {author} {\bibinfo {author} {\bibfnamefont {M.~J.}\ \bibnamefont
		{Dayel}}, \bibinfo {author} {\bibfnamefont {E.~F.~Y.}\ \bibnamefont {Hom}}, \
	and\ \bibinfo {author} {\bibfnamefont {A.~S.}\ \bibnamefont {Verkman}},\
}\href@noop {} {\bibfield  {journal} {\bibinfo  {journal} {Biophys.\ J.}\
}\textbf {\bibinfo {volume} {76}},\ \bibinfo {pages} {2843} (\bibinfo {year}
{1999})}\BibitemShut {NoStop}%
\bibitem [{\citenamefont {Holcman}\ \emph {et~al.}(2018)\citenamefont
	{Holcman}, \citenamefont {Parutto}, \citenamefont {Chambers}, \citenamefont
	{Fantham}, \citenamefont {Young}, \citenamefont {Marciniak}, \citenamefont
	{Kaminski}, \citenamefont {Ron},\ and\ \citenamefont {Avezov}}]{holcman18}%
\BibitemOpen
\bibfield  {author} {\bibinfo {author} {\bibfnamefont {D.}~\bibnamefont
		{Holcman}}, \bibinfo {author} {\bibfnamefont {P.}~\bibnamefont {Parutto}},
	\bibinfo {author} {\bibfnamefont {J.~E.}\ \bibnamefont {Chambers}}, \bibinfo
	{author} {\bibfnamefont {M.}~\bibnamefont {Fantham}}, \bibinfo {author}
	{\bibfnamefont {L.~J.}\ \bibnamefont {Young}}, \bibinfo {author}
	{\bibfnamefont {S.~J.}\ \bibnamefont {Marciniak}}, \bibinfo {author}
	{\bibfnamefont {C.~F.}\ \bibnamefont {Kaminski}}, \bibinfo {author}
	{\bibfnamefont {D.}~\bibnamefont {Ron}}, \ and\ \bibinfo {author}
	{\bibfnamefont {E.}~\bibnamefont {Avezov}},\ }\href@noop {} {\bibfield
	{journal} {\bibinfo  {journal} {Nat.\ Cell Biol.}\ }\textbf {\bibinfo
		{volume} {20}},\ \bibinfo {pages} {1118} (\bibinfo {year}
	{2018})}\BibitemShut {NoStop}%
\bibitem [{\citenamefont {Koslover}\ and\ \citenamefont
	{Spakowitz}(2012)}]{koslover12}%
\BibitemOpen
\bibfield  {author} {\bibinfo {author} {\bibfnamefont {E.~F.}\ \bibnamefont
		{Koslover}}\ and\ \bibinfo {author} {\bibfnamefont {A.~J.}\ \bibnamefont
		{Spakowitz}},\ }\href@noop {} {\bibfield  {journal} {\bibinfo  {journal}
		{Phys.\ Rev.\ E}\ }\textbf {\bibinfo {volume} {86}},\ \bibinfo {pages}
	{011906} (\bibinfo {year} {2012})}\BibitemShut {NoStop}%
\bibitem [{\citenamefont {Maier}\ and\ \citenamefont
	{Brockmann}(2017)}]{maier17}%
\BibitemOpen
\bibfield  {author} {\bibinfo {author} {\bibfnamefont {B.~F.}\ \bibnamefont
		{Maier}}\ and\ \bibinfo {author} {\bibfnamefont {D.}~\bibnamefont
		{Brockmann}},\ }\href@noop {} {\bibfield  {journal} {\bibinfo  {journal}
		{Phys.\ Rev.\ E}\ }\textbf {\bibinfo {volume} {96}},\ \bibinfo {pages}
	{042307} (\bibinfo {year} {2017})}\BibitemShut {NoStop}%
\bibitem [{\citenamefont {Lizana}\ and\ \citenamefont
	{Konkoli}(2005)}]{lizana05}%
\BibitemOpen
\bibfield  {author} {\bibinfo {author} {\bibfnamefont {L.}~\bibnamefont
		{Lizana}}\ and\ \bibinfo {author} {\bibfnamefont {Z.}~\bibnamefont
		{Konkoli}},\ }\href@noop {} {\bibfield  {journal} {\bibinfo  {journal}
		{Phys.\ Rev.\ E}\ }\textbf {\bibinfo {volume} {72}},\ \bibinfo {pages}
	{026305} (\bibinfo {year} {2005})}\BibitemShut {NoStop}%
\bibitem [{\citenamefont {Tejedor}\ \emph {et~al.}(2009)\citenamefont
	{Tejedor}, \citenamefont {Benichou},\ and\ \citenamefont
	{Voituriez}}]{tejedor09}%
\BibitemOpen
\bibfield  {author} {\bibinfo {author} {\bibfnamefont {V.}~\bibnamefont
		{Tejedor}}, \bibinfo {author} {\bibfnamefont {O.}~\bibnamefont {Benichou}}, \
	and\ \bibinfo {author} {\bibfnamefont {R.}~\bibnamefont {Voituriez}},\
}\href@noop {} {\bibfield  {journal} {\bibinfo  {journal} {Phys.\ Rev.\ E}\
}\textbf {\bibinfo {volume} {80}},\ \bibinfo {pages} {065104} (\bibinfo
{year} {2009})}\BibitemShut {NoStop}%
\bibitem [{\citenamefont {Friedman}\ and\ \citenamefont
	{Voeltz}(2011)}]{friedman2011er}%
\BibitemOpen
\bibfield  {author} {\bibinfo {author} {\bibfnamefont {J.~R.}\ \bibnamefont
		{Friedman}}\ and\ \bibinfo {author} {\bibfnamefont {G.~K.}\ \bibnamefont
		{Voeltz}},\ }\href@noop {} {\bibfield  {journal} {\bibinfo  {journal} {Trends
			Cell Biol.}\ }\textbf {\bibinfo {volume} {21}},\ \bibinfo {pages} {709}
	(\bibinfo {year} {2011})}\BibitemShut {NoStop}%
\bibitem [{\citenamefont {Shemesh}\ \emph {et~al.}(2014)\citenamefont
	{Shemesh}, \citenamefont {Klemm}, \citenamefont {Romano}, \citenamefont
	{Wang}, \citenamefont {Vaughan}, \citenamefont {Zhuang}, \citenamefont
	{Tukachinsky}, \citenamefont {Kozlov},\ and\ \citenamefont
	{Rapoport}}]{shemesh14}%
\BibitemOpen
\bibfield  {author} {\bibinfo {author} {\bibfnamefont {T.}~\bibnamefont
		{Shemesh}}, \bibinfo {author} {\bibfnamefont {R.~W.}\ \bibnamefont {Klemm}},
	\bibinfo {author} {\bibfnamefont {F.~B.}\ \bibnamefont {Romano}}, \bibinfo
	{author} {\bibfnamefont {S.}~\bibnamefont {Wang}}, \bibinfo {author}
	{\bibfnamefont {J.}~\bibnamefont {Vaughan}}, \bibinfo {author} {\bibfnamefont
		{X.}~\bibnamefont {Zhuang}}, \bibinfo {author} {\bibfnamefont
		{H.}~\bibnamefont {Tukachinsky}}, \bibinfo {author} {\bibfnamefont {M.~M.}\
		\bibnamefont {Kozlov}}, \ and\ \bibinfo {author} {\bibfnamefont {T.~A.}\
		\bibnamefont {Rapoport}},\ }\href@noop {} {\bibfield  {journal} {\bibinfo
		{journal} {Proc.\ Natl.\ Acad.\ Sci.\ USA}\ }\textbf {\bibinfo {volume}
		{111}},\ \bibinfo {pages} {E5243} (\bibinfo {year} {2014})}\BibitemShut
{NoStop}%
\bibitem [{\citenamefont {Adler}\ \emph {et~al.}(2019)\citenamefont {Adler},
	\citenamefont {Sintorn}, \citenamefont {Strand},\ and\ \citenamefont
	{Parmryd}}]{adler2019conventional}%
\BibitemOpen
\bibfield  {author} {\bibinfo {author} {\bibfnamefont {J.}~\bibnamefont
		{Adler}}, \bibinfo {author} {\bibfnamefont {I.-M.}\ \bibnamefont {Sintorn}},
	\bibinfo {author} {\bibfnamefont {R.}~\bibnamefont {Strand}}, \ and\ \bibinfo
	{author} {\bibfnamefont {I.}~\bibnamefont {Parmryd}},\ }\href@noop {}
{\bibfield  {journal} {\bibinfo  {journal} {Communications biology}\ }\textbf
	{\bibinfo {volume} {2}},\ \bibinfo {pages} {1} (\bibinfo {year}
	{2019})}\BibitemShut {NoStop}%
\bibitem [{\citenamefont {Gefen}\ and\ \citenamefont
	{Aharony}(1983)}]{gefen83}%
\BibitemOpen
\bibfield  {author} {\bibinfo {author} {\bibfnamefont {Y.}~\bibnamefont
		{Gefen}}\ and\ \bibinfo {author} {\bibfnamefont {A.}~\bibnamefont
		{Aharony}},\ }\href@noop {} {\bibfield  {journal} {\bibinfo  {journal} {Phys.
			Rev. Lett.}\ }\textbf {\bibinfo {volume} {50}},\ \bibinfo {pages} {77}
	(\bibinfo {year} {1983})}\BibitemShut {NoStop}%
\bibitem [{\citenamefont {Stauffer}\ and\ \citenamefont
	{Aharony}(1994)}]{stauffer94}%
\BibitemOpen
\bibfield  {author} {\bibinfo {author} {\bibfnamefont {D.}~\bibnamefont
		{Stauffer}}\ and\ \bibinfo {author} {\bibfnamefont {A.}~\bibnamefont
		{Aharony}},\ }\href@noop {} {\emph {\bibinfo {title} {{Introduction to
				percolation theory}}}}\ (\bibinfo  {publisher} {Taylor and Francis},\
\bibinfo {year} {1994})\BibitemShut {NoStop}%
\bibitem [{\citenamefont {Lin}\ \emph {et~al.}(2010)\citenamefont {Lin},
	\citenamefont {Wu},\ and\ \citenamefont {Zhang}}]{lin2010determining}%
\BibitemOpen
\bibfield  {author} {\bibinfo {author} {\bibfnamefont {Y.}~\bibnamefont
		{Lin}}, \bibinfo {author} {\bibfnamefont {B.}~\bibnamefont {Wu}}, \ and\
	\bibinfo {author} {\bibfnamefont {Z.}~\bibnamefont {Zhang}},\ }\href@noop {}
{\bibfield  {journal} {\bibinfo  {journal} {Phys. Rev. E}\ }\textbf {\bibinfo
		{volume} {82}},\ \bibinfo {pages} {031140} (\bibinfo {year}
	{2010})}\BibitemShut {NoStop}%
\bibitem [{\citenamefont {Carretero-Campos}\ \emph {et~al.}(2012)\citenamefont
	{Carretero-Campos}, \citenamefont {Bernaola-Galv{\'a}n}, \citenamefont
	{Ivanov},\ and\ \citenamefont {Carpena}}]{carretero2012phase}%
\BibitemOpen
\bibfield  {author} {\bibinfo {author} {\bibfnamefont {C.}~\bibnamefont
		{Carretero-Campos}}, \bibinfo {author} {\bibfnamefont {P.}~\bibnamefont
		{Bernaola-Galv{\'a}n}}, \bibinfo {author} {\bibfnamefont {P.~C.}\
		\bibnamefont {Ivanov}}, \ and\ \bibinfo {author} {\bibfnamefont
		{P.}~\bibnamefont {Carpena}},\ }\href@noop {} {\bibfield  {journal} {\bibinfo
		{journal} {Physical Review E}\ }\textbf {\bibinfo {volume} {85}},\ \bibinfo
	{pages} {011139} (\bibinfo {year} {2012})}\BibitemShut {NoStop}%
\bibitem [{\citenamefont {Partikian}\ \emph {et~al.}(1998)\citenamefont
	{Partikian}, \citenamefont {{\"O}lveczky}, \citenamefont {Swaminathan},
	\citenamefont {Li},\ and\ \citenamefont {Verkman}}]{partikian1998rapid}%
\BibitemOpen
\bibfield  {author} {\bibinfo {author} {\bibfnamefont {A.}~\bibnamefont
		{Partikian}}, \bibinfo {author} {\bibfnamefont {B.}~\bibnamefont
		{{\"O}lveczky}}, \bibinfo {author} {\bibfnamefont {R.}~\bibnamefont
		{Swaminathan}}, \bibinfo {author} {\bibfnamefont {Y.}~\bibnamefont {Li}}, \
	and\ \bibinfo {author} {\bibfnamefont {A.}~\bibnamefont {Verkman}},\
}\href@noop {} {\bibfield  {journal} {\bibinfo  {journal} {The Journal of
		cell biology}\ }\textbf {\bibinfo {volume} {140}},\ \bibinfo {pages} {821}
(\bibinfo {year} {1998})}\BibitemShut {NoStop}%
\bibitem [{\citenamefont {Sbalzarini}\ \emph {et~al.}(2005)\citenamefont
	{Sbalzarini}, \citenamefont {Mezzacasa}, \citenamefont {Helenius},\ and\
	\citenamefont {Koumoutsakos}}]{sbalzarini2005effects}%
\BibitemOpen
\bibfield  {author} {\bibinfo {author} {\bibfnamefont {I.~F.}\ \bibnamefont
		{Sbalzarini}}, \bibinfo {author} {\bibfnamefont {A.}~\bibnamefont
		{Mezzacasa}}, \bibinfo {author} {\bibfnamefont {A.}~\bibnamefont {Helenius}},
	\ and\ \bibinfo {author} {\bibfnamefont {P.}~\bibnamefont {Koumoutsakos}},\
}\href@noop {} {\bibfield  {journal} {\bibinfo  {journal} {Biophysical
		journal}\ }\textbf {\bibinfo {volume} {89}},\ \bibinfo {pages} {1482}
(\bibinfo {year} {2005})}\BibitemShut {NoStop}%
\bibitem [{\citenamefont {Dieteren}\ \emph {et~al.}(2011)\citenamefont
	{Dieteren}, \citenamefont {Gielen}, \citenamefont {Nijtmans}, \citenamefont
	{Smeitink}, \citenamefont {Swarts}, \citenamefont {Brock}, \citenamefont
	{Willems},\ and\ \citenamefont {Koopman}}]{dieteren2011solute}%
\BibitemOpen
\bibfield  {author} {\bibinfo {author} {\bibfnamefont {C.~E.}\ \bibnamefont
		{Dieteren}}, \bibinfo {author} {\bibfnamefont {S.~C.}\ \bibnamefont
		{Gielen}}, \bibinfo {author} {\bibfnamefont {L.~G.}\ \bibnamefont
		{Nijtmans}}, \bibinfo {author} {\bibfnamefont {J.~A.}\ \bibnamefont
		{Smeitink}}, \bibinfo {author} {\bibfnamefont {H.~G.}\ \bibnamefont
		{Swarts}}, \bibinfo {author} {\bibfnamefont {R.}~\bibnamefont {Brock}},
	\bibinfo {author} {\bibfnamefont {P.~H.}\ \bibnamefont {Willems}}, \ and\
	\bibinfo {author} {\bibfnamefont {W.~J.}\ \bibnamefont {Koopman}},\
}\href@noop {} {\bibfield  {journal} {\bibinfo  {journal} {Proceedings of the
		National Academy of Sciences}\ }\textbf {\bibinfo {volume} {108}},\ \bibinfo
{pages} {8657} (\bibinfo {year} {2011})}\BibitemShut {NoStop}%
\bibitem [{\citenamefont {Yamada}\ \emph {et~al.}(2000)\citenamefont {Yamada},
	\citenamefont {Wirtz},\ and\ \citenamefont {Kuo}}]{yamada2000mechanics}%
\BibitemOpen
\bibfield  {author} {\bibinfo {author} {\bibfnamefont {S.}~\bibnamefont
		{Yamada}}, \bibinfo {author} {\bibfnamefont {D.}~\bibnamefont {Wirtz}}, \
	and\ \bibinfo {author} {\bibfnamefont {S.~C.}\ \bibnamefont {Kuo}},\
}\href@noop {} {\bibfield  {journal} {\bibinfo  {journal} {Biophysical
		journal}\ }\textbf {\bibinfo {volume} {78}},\ \bibinfo {pages} {1736}
(\bibinfo {year} {2000})}\BibitemShut {NoStop}%
\bibitem [{\citenamefont {Toli{\'c}-N{\o}rrelykke}\ \emph
	{et~al.}(2004)\citenamefont {Toli{\'c}-N{\o}rrelykke}, \citenamefont
	{Munteanu}, \citenamefont {Thon}, \citenamefont {Oddershede},\ and\
	\citenamefont {Berg-S{\o}rensen}}]{tolic2004anomalous}%
\BibitemOpen
\bibfield  {author} {\bibinfo {author} {\bibfnamefont {I.~M.}\ \bibnamefont
		{Toli{\'c}-N{\o}rrelykke}}, \bibinfo {author} {\bibfnamefont {E.-L.}\
		\bibnamefont {Munteanu}}, \bibinfo {author} {\bibfnamefont {G.}~\bibnamefont
		{Thon}}, \bibinfo {author} {\bibfnamefont {L.}~\bibnamefont {Oddershede}}, \
	and\ \bibinfo {author} {\bibfnamefont {K.}~\bibnamefont {Berg-S{\o}rensen}},\
}\href@noop {} {\bibfield  {journal} {\bibinfo  {journal} {Phys Rev Lett}\
}\textbf {\bibinfo {volume} {93}},\ \bibinfo {pages} {078102} (\bibinfo
{year} {2004})}\BibitemShut {NoStop}%
\bibitem [{\citenamefont {Lampo}\ \emph {et~al.}(2017)\citenamefont {Lampo},
	\citenamefont {Stylianidou}, \citenamefont {Backlund}, \citenamefont
	{Wiggins},\ and\ \citenamefont {Spakowitz}}]{lampo2017cytoplasmic}%
\BibitemOpen
\bibfield  {author} {\bibinfo {author} {\bibfnamefont {T.~J.}\ \bibnamefont
		{Lampo}}, \bibinfo {author} {\bibfnamefont {S.}~\bibnamefont {Stylianidou}},
	\bibinfo {author} {\bibfnamefont {M.~P.}\ \bibnamefont {Backlund}}, \bibinfo
	{author} {\bibfnamefont {P.~A.}\ \bibnamefont {Wiggins}}, \ and\ \bibinfo
	{author} {\bibfnamefont {A.~J.}\ \bibnamefont {Spakowitz}},\ }\href@noop {}
{\bibfield  {journal} {\bibinfo  {journal} {Biophysical journal}\ }\textbf
	{\bibinfo {volume} {112}},\ \bibinfo {pages} {532} (\bibinfo {year}
	{2017})}\BibitemShut {NoStop}%
\bibitem [{\citenamefont {Etoc}\ \emph {et~al.}(2018)\citenamefont {Etoc},
	\citenamefont {Balloul}, \citenamefont {Vicario}, \citenamefont {Normanno},
	\citenamefont {Li{\ss}e}, \citenamefont {Sittner}, \citenamefont {Piehler},
	\citenamefont {Dahan},\ and\ \citenamefont {Coppey}}]{etoc2018non}%
\BibitemOpen
\bibfield  {author} {\bibinfo {author} {\bibfnamefont {F.}~\bibnamefont
		{Etoc}}, \bibinfo {author} {\bibfnamefont {E.}~\bibnamefont {Balloul}},
	\bibinfo {author} {\bibfnamefont {C.}~\bibnamefont {Vicario}}, \bibinfo
	{author} {\bibfnamefont {D.}~\bibnamefont {Normanno}}, \bibinfo {author}
	{\bibfnamefont {D.}~\bibnamefont {Li{\ss}e}}, \bibinfo {author}
	{\bibfnamefont {A.}~\bibnamefont {Sittner}}, \bibinfo {author} {\bibfnamefont
		{J.}~\bibnamefont {Piehler}}, \bibinfo {author} {\bibfnamefont
		{M.}~\bibnamefont {Dahan}}, \ and\ \bibinfo {author} {\bibfnamefont
		{M.}~\bibnamefont {Coppey}},\ }\href@noop {} {\bibfield  {journal} {\bibinfo
		{journal} {Nature materials}\ }\textbf {\bibinfo {volume} {17}},\ \bibinfo
	{pages} {740} (\bibinfo {year} {2018})}\BibitemShut {NoStop}%
\bibitem [{\citenamefont {Sherman}(1981)}]{sherman81}%
\BibitemOpen
\bibfield  {author} {\bibinfo {author} {\bibfnamefont {T.~F.}\ \bibnamefont
		{Sherman}},\ }\href@noop {} {\bibfield  {journal} {\bibinfo  {journal} {J.\
			Gen.\ Physiol.}\ }\textbf {\bibinfo {volume} {78}},\ \bibinfo {pages} {431}
	(\bibinfo {year} {1981})}\BibitemShut {NoStop}%
\bibitem [{\citenamefont {McCulloh}\ \emph {et~al.}(2003)\citenamefont
	{McCulloh}, \citenamefont {Sperry},\ and\ \citenamefont
	{Adler}}]{mcculloh03}%
\BibitemOpen
\bibfield  {author} {\bibinfo {author} {\bibfnamefont {K.~A.}\ \bibnamefont
		{McCulloh}}, \bibinfo {author} {\bibfnamefont {J.~S.}\ \bibnamefont
		{Sperry}}, \ and\ \bibinfo {author} {\bibfnamefont {F.~R.}\ \bibnamefont
		{Adler}},\ }\href@noop {} {\bibfield  {journal} {\bibinfo  {journal}
		{Nature}\ }\textbf {\bibinfo {volume} {421}},\ \bibinfo {pages} {939}
	(\bibinfo {year} {2003})}\BibitemShut {NoStop}%
\bibitem [{\citenamefont {Tero}\ \emph {et~al.}(2009)\citenamefont {Tero},
	\citenamefont {Takagi}, \citenamefont {Saigusa}, \citenamefont {Ito},
	\citenamefont {Bebber}, \citenamefont {Fricker}, \citenamefont {Yumiki},
	\citenamefont {Kobayashi},\ and\ \citenamefont {Nakagaki}}]{tero10}%
\BibitemOpen
\bibfield  {author} {\bibinfo {author} {\bibfnamefont {A.}~\bibnamefont
		{Tero}}, \bibinfo {author} {\bibfnamefont {S.}~\bibnamefont {Takagi}},
	\bibinfo {author} {\bibfnamefont {T.}~\bibnamefont {Saigusa}}, \bibinfo
	{author} {\bibfnamefont {K.}~\bibnamefont {Ito}}, \bibinfo {author}
	{\bibfnamefont {D.~P.}\ \bibnamefont {Bebber}}, \bibinfo {author}
	{\bibfnamefont {M.~D.}\ \bibnamefont {Fricker}}, \bibinfo {author}
	{\bibfnamefont {K.}~\bibnamefont {Yumiki}}, \bibinfo {author} {\bibfnamefont
		{R.}~\bibnamefont {Kobayashi}}, \ and\ \bibinfo {author} {\bibfnamefont
		{T.}~\bibnamefont {Nakagaki}},\ }\href@noop {} {\bibfield  {journal}
	{\bibinfo  {journal} {Phys.\ Rev.\ E}\ }\textbf {\bibinfo {volume} {80}},\
	\bibinfo {pages} {065104} (\bibinfo {year} {2009})}\BibitemShut {NoStop}%
\bibitem [{\citenamefont {Heaton}\ \emph {et~al.}(2012)\citenamefont {Heaton},
	\citenamefont {Obara}, \citenamefont {Grau}, \citenamefont {Jones},
	\citenamefont {Nakagaki}, \citenamefont {Boddy},\ and\ \citenamefont
	{Fricker}}]{heaton12}%
\BibitemOpen
\bibfield  {author} {\bibinfo {author} {\bibfnamefont {L.}~\bibnamefont
		{Heaton}}, \bibinfo {author} {\bibfnamefont {B.}~\bibnamefont {Obara}},
	\bibinfo {author} {\bibfnamefont {V.}~\bibnamefont {Grau}}, \bibinfo {author}
	{\bibfnamefont {N.}~\bibnamefont {Jones}}, \bibinfo {author} {\bibfnamefont
		{T.}~\bibnamefont {Nakagaki}}, \bibinfo {author} {\bibfnamefont
		{L.}~\bibnamefont {Boddy}}, \ and\ \bibinfo {author} {\bibfnamefont {M.~D.}\
		\bibnamefont {Fricker}},\ }\href@noop {} {\bibfield  {journal} {\bibinfo
		{journal} {Fung.\ Biol.\ Rev.}\ }\textbf {\bibinfo {volume} {26}},\ \bibinfo
	{pages} {12} (\bibinfo {year} {2012})}\BibitemShut {NoStop}%
\bibitem [{\citenamefont {Klecker}\ \emph {et~al.}(2014)\citenamefont
	{Klecker}, \citenamefont {B{\"o}ckler},\ and\ \citenamefont
	{Westermann}}]{klecker2014making}%
\BibitemOpen
\bibfield  {author} {\bibinfo {author} {\bibfnamefont {T.}~\bibnamefont
		{Klecker}}, \bibinfo {author} {\bibfnamefont {S.}~\bibnamefont
		{B{\"o}ckler}}, \ and\ \bibinfo {author} {\bibfnamefont {B.}~\bibnamefont
		{Westermann}},\ }\href@noop {} {\bibfield  {journal} {\bibinfo  {journal}
		{Trends in cell biology}\ }\textbf {\bibinfo {volume} {24}},\ \bibinfo
	{pages} {537} (\bibinfo {year} {2014})}\BibitemShut {NoStop}%
\bibitem [{\citenamefont {Phillips}\ and\ \citenamefont
	{Voeltz}(2016)}]{phillips2016structure}%
\BibitemOpen
\bibfield  {author} {\bibinfo {author} {\bibfnamefont {M.~J.}\ \bibnamefont
		{Phillips}}\ and\ \bibinfo {author} {\bibfnamefont {G.~K.}\ \bibnamefont
		{Voeltz}},\ }\href@noop {} {\bibfield  {journal} {\bibinfo  {journal} {Nature
			reviews Molecular cell biology}\ }\textbf {\bibinfo {volume} {17}},\ \bibinfo
	{pages} {69} (\bibinfo {year} {2016})}\BibitemShut {NoStop}%
\bibitem [{\citenamefont {Valm}\ \emph {et~al.}(2017)\citenamefont {Valm},
	\citenamefont {Cohen}, \citenamefont {Legant}, \citenamefont {Melunis},
	\citenamefont {Hershberg}, \citenamefont {Wait}, \citenamefont {Cohen},
	\citenamefont {Davidson}, \citenamefont {Betzig},\ and\ \citenamefont
	{Lippincott-Schwartz}}]{valm2017applying}%
\BibitemOpen
\bibfield  {author} {\bibinfo {author} {\bibfnamefont {A.~M.}\ \bibnamefont
		{Valm}}, \bibinfo {author} {\bibfnamefont {S.}~\bibnamefont {Cohen}},
	\bibinfo {author} {\bibfnamefont {W.~R.}\ \bibnamefont {Legant}}, \bibinfo
	{author} {\bibfnamefont {J.}~\bibnamefont {Melunis}}, \bibinfo {author}
	{\bibfnamefont {U.}~\bibnamefont {Hershberg}}, \bibinfo {author}
	{\bibfnamefont {E.}~\bibnamefont {Wait}}, \bibinfo {author} {\bibfnamefont
		{A.~R.}\ \bibnamefont {Cohen}}, \bibinfo {author} {\bibfnamefont {M.~W.}\
		\bibnamefont {Davidson}}, \bibinfo {author} {\bibfnamefont {E.}~\bibnamefont
		{Betzig}}, \ and\ \bibinfo {author} {\bibfnamefont {J.}~\bibnamefont
		{Lippincott-Schwartz}},\ }\href@noop {} {\bibfield  {journal} {\bibinfo
		{journal} {Nature}\ }\textbf {\bibinfo {volume} {546}},\ \bibinfo {pages}
	{162} (\bibinfo {year} {2017})}\BibitemShut {NoStop}%
\bibitem [{\citenamefont {Benard}\ and\ \citenamefont
	{Rossignol}(2008)}]{benard2008ultrastructure}%
\BibitemOpen
\bibfield  {author} {\bibinfo {author} {\bibfnamefont {G.}~\bibnamefont
		{Benard}}\ and\ \bibinfo {author} {\bibfnamefont {R.}~\bibnamefont
		{Rossignol}},\ }\href@noop {} {\bibfield  {journal} {\bibinfo  {journal}
		{Antioxidants \& redox signaling}\ }\textbf {\bibinfo {volume} {10}},\
	\bibinfo {pages} {1313} (\bibinfo {year} {2008})}\BibitemShut {NoStop}%
\bibitem [{\citenamefont {Schuck}\ \emph {et~al.}(2009)\citenamefont {Schuck},
	\citenamefont {Prinz}, \citenamefont {Thorn}, \citenamefont {Voss},\ and\
	\citenamefont {Walter}}]{schuck2009membrane}%
\BibitemOpen
\bibfield  {author} {\bibinfo {author} {\bibfnamefont {S.}~\bibnamefont
		{Schuck}}, \bibinfo {author} {\bibfnamefont {W.~A.}\ \bibnamefont {Prinz}},
	\bibinfo {author} {\bibfnamefont {K.~S.}\ \bibnamefont {Thorn}}, \bibinfo
	{author} {\bibfnamefont {C.}~\bibnamefont {Voss}}, \ and\ \bibinfo {author}
	{\bibfnamefont {P.}~\bibnamefont {Walter}},\ }\href@noop {} {\bibfield
	{journal} {\bibinfo  {journal} {Journal of Cell Biology}\ }\textbf {\bibinfo
		{volume} {187}},\ \bibinfo {pages} {525} (\bibinfo {year}
	{2009})}\BibitemShut {NoStop}%
\bibitem [{\citenamefont {Bartsch}\ \emph {et~al.}(2015)\citenamefont
	{Bartsch}, \citenamefont {Liu}, \citenamefont {Bashan},\ and\ \citenamefont
	{Ivanov}}]{bartsch2015network}%
\BibitemOpen
\bibfield  {author} {\bibinfo {author} {\bibfnamefont {R.~P.}\ \bibnamefont
		{Bartsch}}, \bibinfo {author} {\bibfnamefont {K.~K.}\ \bibnamefont {Liu}},
	\bibinfo {author} {\bibfnamefont {A.}~\bibnamefont {Bashan}}, \ and\ \bibinfo
	{author} {\bibfnamefont {P.~C.}\ \bibnamefont {Ivanov}},\ }\href@noop {}
{\bibfield  {journal} {\bibinfo  {journal} {PloS one}\ }\textbf {\bibinfo
		{volume} {10}} (\bibinfo {year} {2015})}\BibitemShut {NoStop}%
\bibitem [{\citenamefont {Banavar}\ \emph {et~al.}(2000)\citenamefont
	{Banavar}, \citenamefont {Colaiori}, \citenamefont {Flammini}, \citenamefont
	{Maritan},\ and\ \citenamefont {Rinaldo}}]{banavar00}%
\BibitemOpen
\bibfield  {author} {\bibinfo {author} {\bibfnamefont {J.~R.}\ \bibnamefont
		{Banavar}}, \bibinfo {author} {\bibfnamefont {F.}~\bibnamefont {Colaiori}},
	\bibinfo {author} {\bibfnamefont {A.}~\bibnamefont {Flammini}}, \bibinfo
	{author} {\bibfnamefont {A.}~\bibnamefont {Maritan}}, \ and\ \bibinfo
	{author} {\bibfnamefont {A.}~\bibnamefont {Rinaldo}},\ }\href@noop {}
{\bibfield  {journal} {\bibinfo  {journal} {Phys.\ Rev.\ Lett.}\ }\textbf
	{\bibinfo {volume} {84}},\ \bibinfo {pages} {4745} (\bibinfo {year}
	{2000})}\BibitemShut {NoStop}%
\bibitem [{\citenamefont {Bohn}\ and\ \citenamefont {Magnasco}(2007)}]{bohn07}%
\BibitemOpen
\bibfield  {author} {\bibinfo {author} {\bibfnamefont {S.}~\bibnamefont
		{Bohn}}\ and\ \bibinfo {author} {\bibfnamefont {M.~O.}\ \bibnamefont
		{Magnasco}},\ }\href@noop {} {\bibfield  {journal} {\bibinfo  {journal}
		{Phys.\ Rev.\ Lett.}\ }\textbf {\bibinfo {volume} {98}},\ \bibinfo {pages}
	{088702} (\bibinfo {year} {2007})}\BibitemShut {NoStop}%
\bibitem [{\citenamefont {Bernot}\ \emph {et~al.}(2009)\citenamefont {Bernot},
	\citenamefont {Caselles},\ and\ \citenamefont {Morel}}]{bernot09}%
\BibitemOpen
\bibfield  {author} {\bibinfo {author} {\bibfnamefont {M.}~\bibnamefont
		{Bernot}}, \bibinfo {author} {\bibfnamefont {V.}~\bibnamefont {Caselles}}, \
	and\ \bibinfo {author} {\bibfnamefont {J.-M.}\ \bibnamefont {Morel}},\
}\href@noop {} {\emph {\bibinfo {title} {{Optimal Transportation
			Networks}}}}\ (\bibinfo  {publisher} {Springer},\ \bibinfo {address}
{Berlin/Heidelberg},\ \bibinfo {year} {2009})\BibitemShut {NoStop}%
\bibitem [{\citenamefont {Hu}\ and\ \citenamefont {Cai}(2013)}]{hu13}%
\BibitemOpen
\bibfield  {author} {\bibinfo {author} {\bibfnamefont {D.}~\bibnamefont
		{Hu}}\ and\ \bibinfo {author} {\bibfnamefont {D.}~\bibnamefont {Cai}},\
}\href@noop {} {\bibfield  {journal} {\bibinfo  {journal} {Phys.\ Rev.\
		Lett.}\ }\textbf {\bibinfo {volume} {111}},\ \bibinfo {pages} {138701}
(\bibinfo {year} {2013})}\BibitemShut {NoStop}%
\bibitem [{\citenamefont {Katifori}\ \emph {et~al.}(2010)\citenamefont
	{Katifori}, \citenamefont {Szollosi},\ and\ \citenamefont
	{Magnasco}}]{katifori10}%
\BibitemOpen
\bibfield  {author} {\bibinfo {author} {\bibfnamefont {E.}~\bibnamefont
		{Katifori}}, \bibinfo {author} {\bibfnamefont {G.~J.}\ \bibnamefont
		{Szollosi}}, \ and\ \bibinfo {author} {\bibfnamefont {M.~O.}\ \bibnamefont
		{Magnasco}},\ }\href@noop {} {\bibfield  {journal} {\bibinfo  {journal}
		{Phys.\ Rev.\ Lett.}\ }\textbf {\bibinfo {volume} {104}},\ \bibinfo {pages}
	{048704} (\bibinfo {year} {2010})}\BibitemShut {NoStop}%
\bibitem [{\citenamefont {Corson}(2010)}]{corson10}%
\BibitemOpen
\bibfield  {author} {\bibinfo {author} {\bibfnamefont {F.}~\bibnamefont
		{Corson}},\ }\href@noop {} {\bibfield  {journal} {\bibinfo  {journal} {Phys.\
			Rev.\ Lett.}\ }\textbf {\bibinfo {volume} {104}},\ \bibinfo {pages} {048703}
	(\bibinfo {year} {2010})}\BibitemShut {NoStop}%
\bibitem [{\citenamefont {Hoyer}\ \emph {et~al.}(2018)\citenamefont {Hoyer},
	\citenamefont {Chitwood}, \citenamefont {Ebmeier}, \citenamefont {Striepen},
	\citenamefont {Qi}, \citenamefont {Old},\ and\ \citenamefont
	{Voeltz}}]{hoyer18}%
\BibitemOpen
\bibfield  {author} {\bibinfo {author} {\bibfnamefont {M.~J.}\ \bibnamefont
		{Hoyer}}, \bibinfo {author} {\bibfnamefont {P.~J.}\ \bibnamefont {Chitwood}},
	\bibinfo {author} {\bibfnamefont {C.~C.}\ \bibnamefont {Ebmeier}}, \bibinfo
	{author} {\bibfnamefont {J.~F.}\ \bibnamefont {Striepen}}, \bibinfo {author}
	{\bibfnamefont {R.~Z.}\ \bibnamefont {Qi}}, \bibinfo {author} {\bibfnamefont
		{W.~M.}\ \bibnamefont {Old}}, \ and\ \bibinfo {author} {\bibfnamefont
		{G.~K.}\ \bibnamefont {Voeltz}},\ }\href@noop {} {\bibfield  {journal}
	{\bibinfo  {journal} {Cell}\ }\textbf {\bibinfo {volume} {175}},\ \bibinfo
	{pages} {254} (\bibinfo {year} {2018})}\BibitemShut {NoStop}%
\bibitem [{\citenamefont {English}\ and\ \citenamefont
	{Voeltz}(2013{\natexlab{b}})}]{english13b}%
\BibitemOpen
\bibfield  {author} {\bibinfo {author} {\bibfnamefont {A.~R.}\ \bibnamefont
		{English}}\ and\ \bibinfo {author} {\bibfnamefont {G.~K.}\ \bibnamefont
		{Voeltz}},\ }\href@noop {} {\bibfield  {journal} {\bibinfo  {journal} {Nat.
			Cell Biol.}\ }\textbf {\bibinfo {volume} {15}},\ \bibinfo {pages} {169}
	(\bibinfo {year} {2013}{\natexlab{b}})}\BibitemShut {NoStop}%
\bibitem [{\citenamefont {Schindelin}\ \emph {et~al.}(2012)\citenamefont
	{Schindelin}, \citenamefont {Arganda-Carreras}, \citenamefont {Frise},
	\citenamefont {Kaynig}, \citenamefont {Longair}, \citenamefont {Pietzsch},
	\citenamefont {Preibisch}, \citenamefont {Rueden}, \citenamefont {Saalfeld},
	\citenamefont {Schmid}, \citenamefont {Tinevez}, \citenamefont {White},
	\citenamefont {Hartenstein}, \citenamefont {Eliceiri}, \citenamefont
	{Tomancak},\ and\ \citenamefont {Cardona}}]{schindelin12}%
\BibitemOpen
\bibfield  {author} {\bibinfo {author} {\bibfnamefont {J.}~\bibnamefont
		{Schindelin}}, \bibinfo {author} {\bibfnamefont {I.}~\bibnamefont
		{Arganda-Carreras}}, \bibinfo {author} {\bibfnamefont {E.}~\bibnamefont
		{Frise}}, \bibinfo {author} {\bibfnamefont {V.}~\bibnamefont {Kaynig}},
	\bibinfo {author} {\bibfnamefont {M.}~\bibnamefont {Longair}}, \bibinfo
	{author} {\bibfnamefont {T.}~\bibnamefont {Pietzsch}}, \bibinfo {author}
	{\bibfnamefont {S.}~\bibnamefont {Preibisch}}, \bibinfo {author}
	{\bibfnamefont {C.}~\bibnamefont {Rueden}}, \bibinfo {author} {\bibfnamefont
		{S.}~\bibnamefont {Saalfeld}}, \bibinfo {author} {\bibfnamefont
		{B.}~\bibnamefont {Schmid}}, \bibinfo {author} {\bibfnamefont {J.-Y.}\
		\bibnamefont {Tinevez}}, \bibinfo {author} {\bibfnamefont {D.~J.}\
		\bibnamefont {White}}, \bibinfo {author} {\bibfnamefont {V.}~\bibnamefont
		{Hartenstein}}, \bibinfo {author} {\bibfnamefont {K.}~\bibnamefont
		{Eliceiri}}, \bibinfo {author} {\bibfnamefont {P.}~\bibnamefont {Tomancak}},
	\ and\ \bibinfo {author} {\bibfnamefont {A.}~\bibnamefont {Cardona}},\
}\href@noop {} {\bibfield  {journal} {\bibinfo  {journal} {Nature Methods}\
}\textbf {\bibinfo {volume} {9}},\ \bibinfo {pages} {676} (\bibinfo {year}
{2012})}\BibitemShut {NoStop}%
\bibitem [{\citenamefont {Condamin}\ \emph {et~al.}(2008)\citenamefont
	{Condamin}, \citenamefont {Tejedor}, \citenamefont {Voituriez}, \citenamefont
	{Benichou},\ and\ \citenamefont {Klafter}}]{condamin08}%
\BibitemOpen
\bibfield  {author} {\bibinfo {author} {\bibfnamefont {S.}~\bibnamefont
		{Condamin}}, \bibinfo {author} {\bibfnamefont {V.}~\bibnamefont {Tejedor}},
	\bibinfo {author} {\bibfnamefont {R.}~\bibnamefont {Voituriez}}, \bibinfo
	{author} {\bibfnamefont {O.}~\bibnamefont {Benichou}}, \ and\ \bibinfo
	{author} {\bibfnamefont {J.}~\bibnamefont {Klafter}},\ }\href@noop {}
{\bibfield  {journal} {\bibinfo  {journal} {Proc.\ Nat.\ Acad.\ Sci.}\
	}\textbf {\bibinfo {volume} {105}},\ \bibinfo {pages} {5675} (\bibinfo {year}
	{2008})}\BibitemShut {NoStop}%
\end{thebibliography}

%

\end{document}